\documentclass[aps,twocolumn,prb]{revtex4-1}
\usepackage{longtable}
\usepackage{subfigure}
\usepackage{amsmath,amssymb,amsfonts,bm}
\usepackage{graphicx}
\usepackage{ulem}
\usepackage{color}

\def\KeyWord#1{$\backslash$\IfColor{$\!\!$\textRed{#1}\textBlack}{#1}$\!\!$}
\newcommand{\be}{\begin{equation} }
\newcommand{\ee}{\end{equation} }
\newcommand{\ba}{\begin{eqnarray} }
\newcommand{\ea}{\end{eqnarray} }
\newcommand{\n}{\nonumber \\ }

\newcommand{\bit}{\begin{itemize}}
\newcommand{\eit}{\end{itemize}}

\graphicspath{{Figures/}}

\begin{document}
\title{Measuring fermion parity correlations and relaxation rates in 1D topological superconducting wires}
\author{F. J.  Burnell}
\affiliation{All Souls College, Oxford, OX14AL, UK}
\affiliation{Rudolf Peirels center for theoretical physics, Oxford, UK}
\author{Alexander Shnirman}
\affiliation{Institute for Theory of Condensed Matter and DFG Center for Functional Nanostructures (CFN), Karlsruhe Institute of Technology (KIT), Karlsruhe, Germany}
\affiliation{Department of Condensed Matter Physics, Weizmann Institute
of Science, Rehovot, 76100, Israel}
\author{Yuval Oreg}
\affiliation{Department of Condensed Matter Physics, Weizmann Institute
of Science, Rehovot, 76100, Israel}

\begin{abstract}
%Zero energy Majorana fermion states (Majoranas) may be formed in semiconductor wire that is in proximity to a superconductor. For  a successful operation of topological quantum gates based on Majoranas  the parity of the number of particles has to be preserved for a {\it parity time} longer than the gate operation time. In this Letter we suggest a detail protocol for measuring the parity time of a system build of four Majoranas. The suggested protocol requires measurement of temporal charge correlations on segments of the wire at frequencies of order of the parity rate.
%
Zero energy Majorana fermion states (Majoranas) can arise at the ends of a semiconducting wire in proximity with a superconductor. A first generation of experiments has detected a zero bias conductance peak in these systems that strongly suggests these Majoranas do exist; however, a definitive demonstration of the long-ranged entanglement that is crucial for potential applications in quantum computing has yet to be carried out.   This work discusses two possible measurement schemes to detect this long-ranged entanglement  in a wire system with two coupled pairs of Majoranas, by varying the coupling between one pair while measuring the fermion parity of the second pair. First, in a system with two coupled pairs of Majoranas, we discuss how varying the coupling of one pair in time, while measuring temporal fermion parity correlations of the second pair, allows for an experimental probe of long-ranged Majorana entanglement.  Second, we show that the power spectrum of the charge noise (fermion parity noise) of one pair contains signatures of these correlations, as well as allowing one to infer the parity relaxation rate.    
\end{abstract}
\maketitle

The suggestion~\cite{KitaevToric,KitaevMajorana} that topologically protected states of matter could be used for fault-tolerant quantum computation has lead to a flurry of research with the aim of realizing such systems in nature.  Zero energy Majorana fermion states (Majoranas) have caught researchers' interest and imagination as they are (in theory) relatively simple to realize physically: they can arise as bound states in  vortex cores of a topological spinless p-wave superconductor~\cite{ReadGreen}.   As such, they constitute the first (potentially) physically realizable example of non-abelian quasi-particles: exchanging two Majorana zero modes leads to a topologically protected evolution of the system between its degenerate ground states.  Because of their non-abelian statistics, Majoranas can be utilized as fault-tolerant quantum bits for (non-universal) quantum computation.

The possibility of fabricating systems that support Majoranas~\cite{LutchynPRL105.077001, OregPRL105.177002} in semiconducting wires (in proximity to superconductors) and the experimental observation of zero-energy states at the endpoints of these wires~\cite{MourikScience336,DasNatPhys8,Churchill2013,DengNanoLett12} has attracted further interest in these systems. In these topological $p$-wave spinless superconducting wires it can be shown that Majoranas appear on the boundary between  topological and non-topological superconducting (or insulating) regions of the wire.
Hence a Majorana zero mode exists at each end of a topological superconducting wire segment.  Any such {\it pair} of Majoranas can collectively be in one of two states, identified by their fermion parity $n_f =0,1$.

The basis of a quantum qubit constructed from Majoranas is this degeneracy between states of even ($n_f =0$) and odd ($n_f =1$) fermion parity, and the fact that the fermion parities of pairs of Majoranas can be entangled.
Demonstrating the existence of such entanglement in superconducting wires is therefore of fundamental importance. % to any applications in quantum information.

In this work, we present a measurement protocol, which we call the {\it parity experiment}, that can be used to reveal correlations between pairs of the bound states responsible for the experimentally observed $0$-bias peaks, %probing the nature of the zero-energy excitations beyond what can be achieved in tunneling experiments.  
and demonstrate the existence of the fermion parity entanglement that is a necessary prerequisite to any successful applications in quantum information.
Though not as decisive as an actual braiding experiment~\cite{AliceaNatPhys7,SauPRB84}, the parity experiment probes distictively Majorana-like long ranged entanglement between different wire segments, and has the advantage of being significantly simpler to carry out.

The limiting factor in using Majoranas as quantum bits is expected to be the time-scale over which the total fermion parity of a system with many Majoranas remains constant. We refer to this timescale as the {\it parity time}  $\Gamma^{-1}$.  In addition to the parity experiment mentioned above, we will also discuss how to measure the noise spectrum -- and hence the parity time --  with our set-up.   In part, this is useful since $\Gamma^{-1}$ is difficult to predict theoretically. (See Refs.~\cite{Fu09,LutchynPRL105.077001} for alternative methods of measuring $\Gamma^{-1}$.)  Interestingly, however, in our set-up the noise spectrum is also expected to show signatures of correlations between pairs of Majoranas.  Because measuring the noise spectrum is potentially simpler than carrying out the parity experiment itself, this may be the easiest way to search for evidence of long-ranged entanglement in superconducting nanowire systems.

 \begin{figure}[h]
 \includegraphics[width=1.0\linewidth]{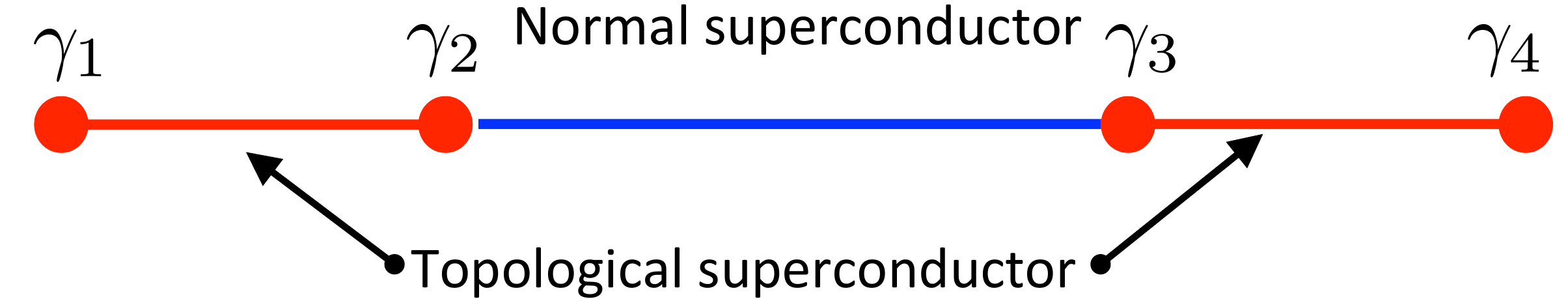}
 %\\
 %\includegraphics[width=0.4\linewidth]{WireFigb.pdf}
 \caption{ Experimental set-up for 1D superconducting wires required for  the parity experiment.  Red lines indicate segments of the wire that are in the topological superconducting state; blue lies indicate segments in the normal superconducting state.  The red dot at the boundary between normal and topological superconducting regions indicates the potential Majorana zero mode.  The auto-correlation function $\langle n_R (t+\tau) n_R(t) \rangle$ of the fermion parity in the right superconducting segment is measured while the coupling $t_{12}$ between the two Majoranas in the left segment of the wire oscillates in time to reveal correlations between left- and right- wire segments due to conservation of the overall fermion parity. }
\label{WireFig}
 \end{figure}

The experimental set-up is shown in Fig. \ref{WireFig}.  It requires a wire with two segments in the topological superconducting regime, separated by a middle segment in the normal superconducting regime.  The experiment consists of measuring the fermion parity of the right wire segment $n_R(t)$ as the chemical potential in the left wire segment oscillates at a frequency $\omega_0$.  We show that for realistic values~\cite{RainisLossPRB85.174533} of $\Gamma^{-1}$, if the 0-bias peaks are truly signatures of Majorana fermions, this will result in oscillations of the fermion parity of the right wire segment.  These oscillations are due to the approximate conservation of the {\it net} fermion parity of both wire segments on the time-scales at which the measurement is performed.  The observation of such correlations would be strongly suggestive that the 0-bias peaks are fermionic in origin, and hence very likely to be Majorana zero modes.

%Before discussing the details of our proposal, let us comment briefly on its viability.
The principle experimental advance required to carry out our proposal is the ability to measure the local fermion parity in one wire segment.  A number of theoretical proposals for making such a measurement exist
\cite{BondersonPRL106,HasslerNJP12,HasslerNJP13}.  In our set-up, we can exploit the fact that when the coupling between two Majoranas is non zero their parity can be detected by a nearby quantum dot which measures the potential created by an additional electron in the wire~\cite{Ben-Shach2013}.
%In addition to the parity measurement, we require a time-varying gate voltage to tune the coupling between the two Majorana fermions on the left-hand wire segment.

\begin{figure}[h]
 \begin{tabular}{ll}
 (a)& \\
 & \includegraphics[width=0.9\linewidth]{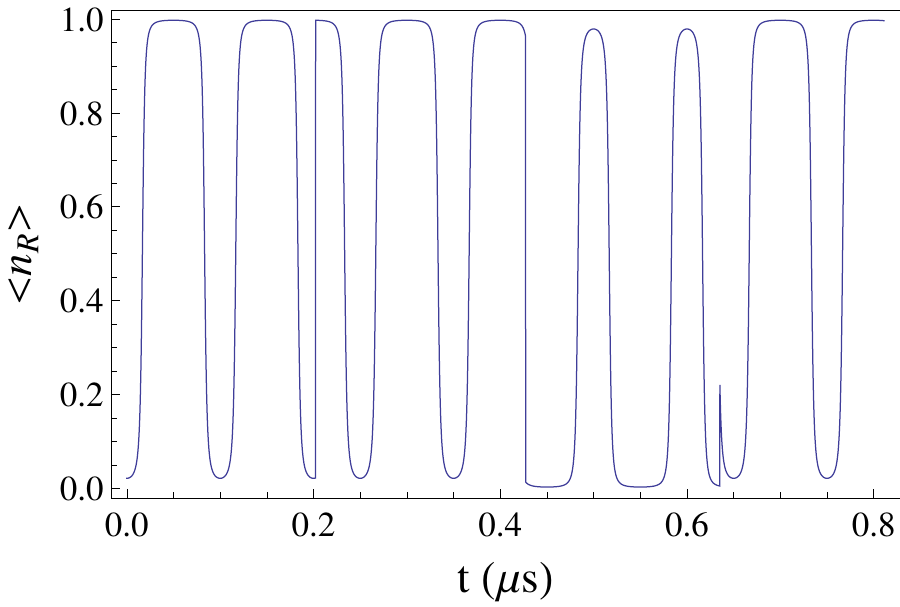} \\
  (b)& \\
  & \includegraphics[width=0.9\linewidth]{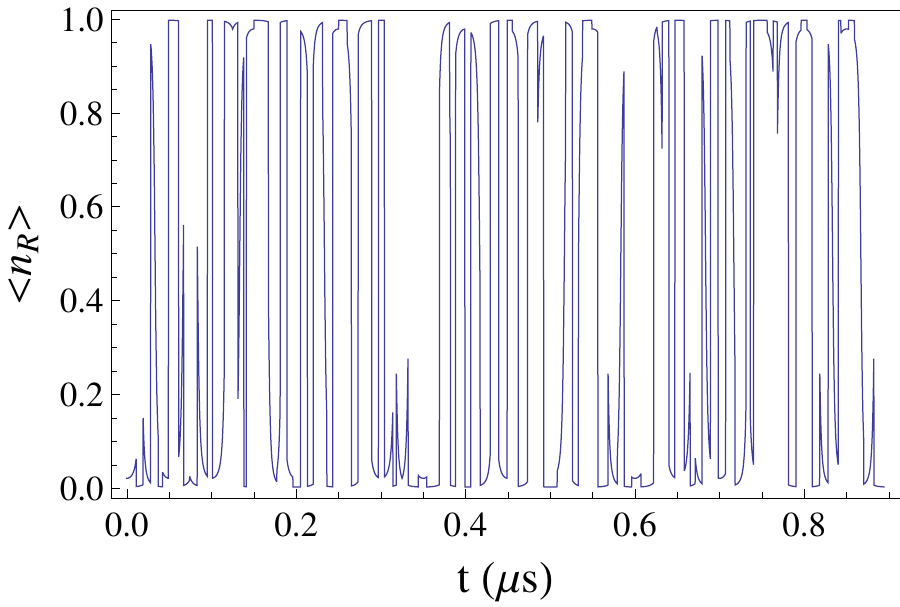}
   \end{tabular}
 \caption{Sample outcome for the parity experiment if (a) the parity time $\Gamma^{-1}$ is longer than the period of oscillation $\omega_0/2\pi$ ($\Gamma=5 MHz, \omega_0/(2 \pi) = 10 MHz$), and (b) the parity time is shorter than the period of oscillation ($\Gamma=120 MHz, \omega_0/(2 \pi) = 10 MHz$). }
\label{fg:ParityTimeFig}
 \end{figure}

An example of the expected time-dependence of the measured fermion parity is shown in Fig. \ref{fg:ParityTimeFig}.  If the parity relaxation rate $\Gamma$ (i.e. the inverse of the parity time)  is slow relative to the driving frequency, the fermion parity of the right wire segment shows clear oscillations in time, in spite of the fact that the chemical potential is changing only in the left wire segment.  Specifically, a continuous measurement of the parity will show oscillations interspersed by jumps, which are the result of stochastic processes that do not conserve the total fermion parity in the wire (Fig \ref{fg:ParityTimeFig} a).  As the relaxation rate increases (relative to the driving frequency), the jumps become more frequent and eventually overwhelm the oscillatory signal (Fig \ref{fg:ParityTimeFig} b).   We describe in detail how these curves were obtained in Sect. \ref{ParityMeas}, after having introduced the basic features of our model.

The remainder of this work is structured as follows.  In Sect. \ref{ExpSec}, we present a model Hamiltonian that describes the dominant interactions between our four Majoranas, and discuss how its spectrum allows us to infer non-local correlations.  Sect. \ref{RelaxSect} discusses how to couple this system to a fermionic bath, and present the equations governing the dynamics of the Majoranas in the presence of such a coupling.  Some of the details are relegated to Appendix \ref{MasterApp}.  In Sect. \ref{ParityMeas} we present the theoretical outcome of our parity experiment, in a parameter regime that is expected to be relevant to current experiments.  Some additional results are presented in Appendix \ref{LowTApp}.  Finally, in Sect. \ref{NoiseSec}, we describe signatures of Majorana correlations that can be observed in the noise spectrum (i.e. by measuring the system's relaxation time, with all couplings fixed) using our set-up.

\section{The parity experiment}\label{ExpSec}

We now describe the details of the parity experiment.  We begin with the Hamiltonian for a completely isolated wire system, at energy scales much lower than the induced superconducting gap in the wire.  In this regime the only degrees of freedom are the Majorana zero modes at the ends of each topological wire segment (cf. Fig~\ref{WireFig}).  Keeping only couplings between neighboring Majoranas, the effective Hamiltonian is:
\be
H = i t_{12} \gamma_1 \gamma_2 + i t_{23} \gamma_2 \gamma_3 + i t_{34} \gamma_3 \gamma_4  \ \ \ .
\ee
 By tuning the chemical potential, wire length, or Zeeman field  (or all of them together) in the left wire segment\cite{DasSarmaSmoking,RainisPRBB87}, we  make $t_{12}$ vary in time according to
\be
t_{12} = t_{12}^{(0)} \cos \omega_0 t
\ee
while keeping $t_{23}, t_{34}$ fixed.

It is convenient to re-express this Hamiltonian in terms of the fermion creation and anhiliation operators associated with the fermion parity of each topological wire segment:
\ba
c^\dag_L =& \frac{1}{2} \left(  \gamma_1 + i \gamma_2 \right), \ \ \ \ c^\dag_R =& \frac{1}{2} \left(  \gamma_3 + i \gamma_4 \right),  \n
c_L =&\frac{1}{2}  \left(  \gamma_1 - i \gamma_2 \right), \ \ \ \ c_R =& \frac{1}{2}  \left(  \gamma_3 - i \gamma_4 \right).
\ea
We may now express our Hamiltonian in the basis of the two fermion numbers $n_L$ and $n_R$,
\be
|n_L, n_R \rangle =(  |0,0 \rangle, |1,1 \rangle |0,1 \rangle, |1,0 \rangle)^T
\ee
In this basis, the matrix elements of the Hamiltonian $H$ are:
\be \label{HEqn}
%\langle n_L, n_R|
 \begin{pmatrix} t_{12} + t_{34} & t_{23} & 0 & 0 \\ t_{23} & -t_{12}  - t_{34} & 0 & 0 \\  0 & 0 & t_{12} - t_{34} & t_{23} \\
0 & 0& t_{23} & - t_{12} + t_{34}   \\
\end{pmatrix}.
%|n_L, n_R \rangle
\ee

\begin{figure}
% \begin{tabular}{cc}
% \includegraphics[width=0.4\linewidth]{Fig1a.pdf} & \includegraphics[width=0.4\linewidth]{Fig1b.pdf} \\
 %\end{tabular}
 \begin{tabular}{ll}
 (a)& \\
 & \includegraphics[width=0.8\linewidth]{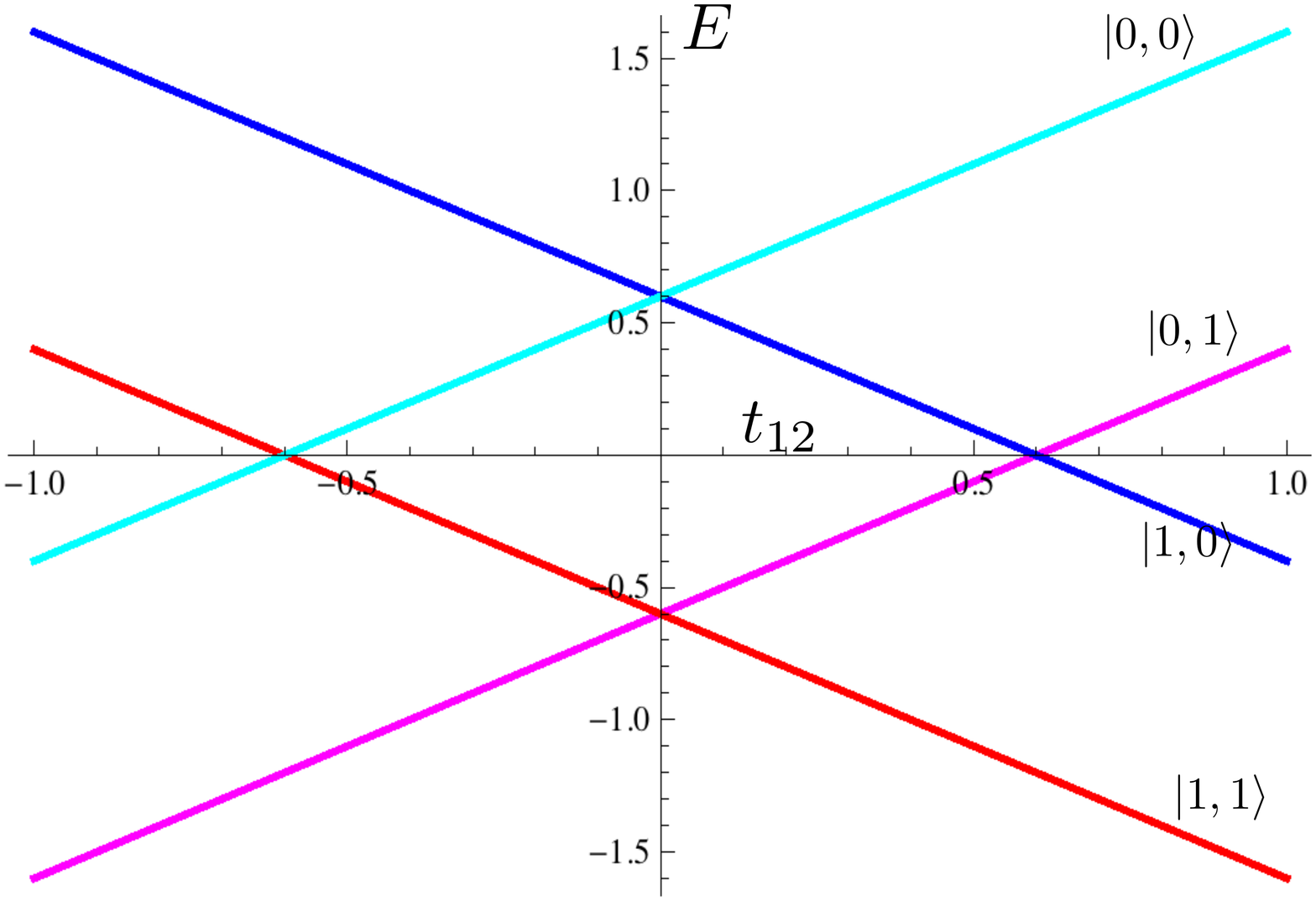} \\
  (b)& \\
  & \includegraphics[width=0.8\linewidth]{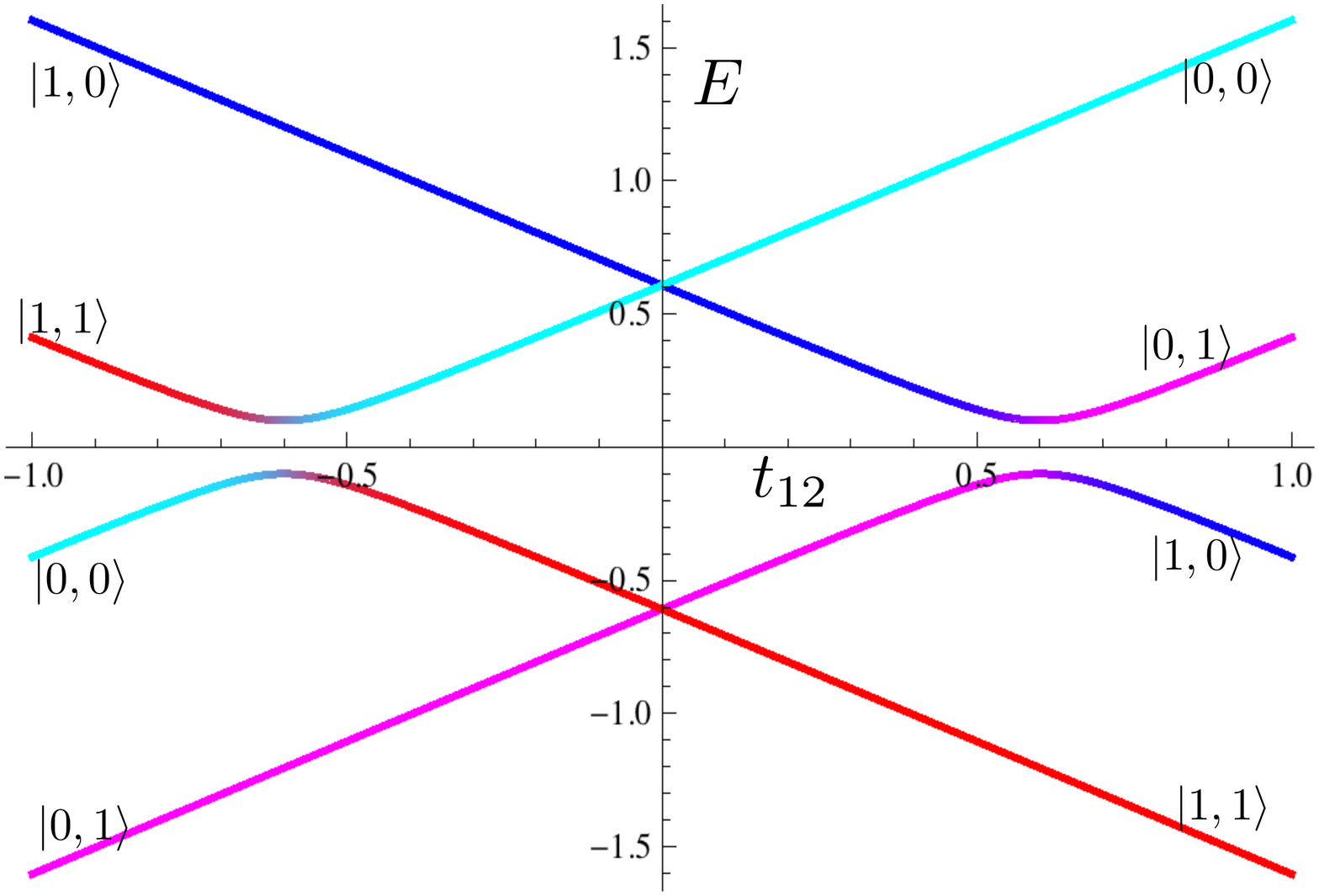}
   \end{tabular}
 \caption{ %{\bf Yuval: and the states on the left part of the figure as well, make the line bolder, add labels for x axis and y axis} 
 Band structure and weights of relative states for $t_{34} =0.6$ and $t_{23} =0$ (a), $t_{23} = 0.2$ (b), shown for $t_{12}$ ranging from $-1$ to $1$.  Coloring shows the relative weights of the 4 possible eigenstates in these bands. Turquoise is $|0,0 \rangle$, red is $|1,1 \rangle$, purple is $| 0,1 \rangle$, and  blue is $|1,0 \rangle$. }
\label{Fig1}
 \end{figure}

Fig. \ref{Fig1} shows the band structure of this model for fixed $t_{23}< t_{34}<t_{12}^{(0)}$, and for $t_{12}$ scanning from $t_{12}^{(0)}$ to $-t_{12}^{(0)}$.
For $t_{23} =0$ (Fig. \ref{Fig1}a), in each fermion parity sector there is a single crossing, which occurs at $t_{12} =- t_{34} $ ($t_{12} = t_{34} $) in the even (odd) sector.  For $t_{23} >0$ (Fig. \ref{Fig1}b) these crossings are avoided, separated by an energy gap of $2 t_{23}$.  The colors indicate the dominant fermion number composition of each eigenstate: turquoise is $|0,0 \rangle$, red is $|1,1 \rangle$, purple is $| 0,1 \rangle$, and  blue is $|1,0 \rangle$. For $t_{23}$ small the states of fixed $n_L, n_R$ are reasonable approximations to the eigenstates, except near the avoided crossings.

In addition to the avoided crossing between states in the same fermion parity sector, there is always an unavoided crossing between states in sectors of different fermion parity at $t_{12} =0$, where the states $|0,0 \rangle, |1,0 \rangle$ are degenerate, as are $|1,1 \rangle, |0,1 \rangle$.  This crossing is protected by the conservation of total fermion parity, and can only be lifted by processes in which an odd number of fermions tunnel into or out of the superconducting wire system.  (We will discuss such processes in Sec. \ref{RelaxSect}).
%Hence if we start in the ground state with $t_{12} >0$ (which is $|0,0 \rangle$), then once $t_{12} <0$ we are in an excited state.

 The principle of our proposed experimental protocol  is as follows: we begin with $t_{12} > t_{34}>0$, such that the ground state of the system is $|1,1 \rangle$ (red in Fig.~\ref{Fig1}), with energy approximately $-t_{12} - t_{34}$. (Here we are neglecting corrections of order $t_{23}/t_{12}, t_{23}/t_{34}$ in both the energies and the composition of the eigenstates).  At this point the next lowest state in energy is $|1,0 \rangle$ ( $E=-t_{12} + t_{34}$; blue), followed by $|0,1 \rangle$ ( $E= t_{12} - t_{34}$; purple) and $|0,0 \rangle$ ( $E= t_{12} + t_{34}$; turquoise).  We then vary $t_{12}$ to a final value where $t_{12} + t_{34} <0$.  Here the ground state is $|0,1 \rangle$ (purple), and the next lowest state in energy is $|0,0 \rangle$ (turquoise), followed by $|1,1 \rangle$ and $|1,0 \rangle$.   However, if the total fermion parity of the system is conserved, it cannot relax to its true ground state, which is in a sector of different fermion parity.
  Thus if we vary $t_{12}$ slowly (relative to $2 t_{23}$, which sets the gap at the avoided crossing),
 %that the process is adiabatic about the avoided crossing,% between $|0,0 \rangle$ and $|1,1 \rangle$,
 the system will remain in the lowest- energy eigenstate {\it of total even fermion parity}, which changes from being predominantly $|1,1 \rangle$ to predominantly  $|0,0 \rangle$.  Experimentally, the result is that although $t_{34}$ remains fixed, the fermion number on the right end of the wire changes from $1$ to $0$ as we vary $t_{12}$, giving a distinctive signature of the long-ranged entanglement of the Majorana fermions.

\section{Time evolution of the density matrix with relaxation processes} \label{RelaxSect}
%Let us consider how this effect can manifest itself in a more realistic experimental system.
  In practice, it is not possible to isolate the wire such that the fermion parity is strictly conserved; both the coupling to external leads and quasi-particles in the bulk superconductor can spoil fermion parity conservation.  The system will therefore have a finite parity relaxation time $\Gamma^{-1}$ after which the parity-even first-excited state at $t_{12} = - t_{12}^{(0)}$ will relax to the parity-odd ground state.
To model such relaxation processes, we  couple our wire system to  a fermionic bath: %fermion operators $f^\dag_{L, R}$ external to the system:
\ba\label{eq:deltaH}
\delta H &=& \sum_\epsilon \left[\alpha_R^{(1)} c^\dag_R f_{R,\epsilon} +  \overline{\alpha}_R^{(1)} f^\dag_{R,\epsilon} c_R + \alpha_R^{(2)} c^\dag_R f^\dag_{R,\epsilon} +  \overline{\alpha}_R^{(2)}  f_{R,\epsilon} c_R \right]   \n
&&
+\sum_\varepsilon\left[ \alpha_L^{(1)}  c^\dag_L f_{L,\varepsilon} + \overline{\alpha}_L^{(1)} f^\dag_{L,\varepsilon} c_L
+ \alpha_L^{(2)}   c^\dag_L f^\dag_{L,\varepsilon} +  \overline{\alpha}_L^{(2)} f^\dag_{L,\varepsilon} c_L \right] \n
&&
+ H^{\text{Bath}} (f^\dag_L, f_L, f^\dag_R, f_R)
\ea
Here the indices $\epsilon$ and $\varepsilon$ run over the continuum of states in the left and right reservoirs respectively, with $f^\dag_{L, \epsilon} f^\dag_{ R,\varepsilon}$ the corresponding fermion creation operators.
For simplicity we assume the tunneling amplitudes $\alpha^{(i)}_{L,R}$ to be $\epsilon$ and $\varepsilon$ independent.

A relaxation transition in which the energy of the fermion modes in the wire changes by $- \Delta E$ creates an excitation (particle-like or hole-like) of energy $ \Delta E$ in the bath.
We will take the bath to be at equilibrium at some temperature $T$, so that the probability of creating an excitation of energy $E$ in the bath is:
\be
\langle f_{R,E}^\dag f_{R,E} \rangle = n_F(E -\mu_L) \quad,\quad \langle f_{L,E}^\dag f_{L,E} \rangle = n_F(E -\mu_R)\ .
\ee
where $n_F (E) = 1/( 1+ e^{ E/ (K_B T)} )$.

The dynamics of the density matrix $\rho$ of our wire system coupled to the
fermionic bath is described by:
\be \label{MasterEq}
\dot{\rho} =- \frac{i}{\hbar} \left[ H, \rho \right] + \sum_n \,\Gamma_n\,\left[ L_n \rho L^\dag_n - \frac{1}{2} \left( L^\dag_n L_n \rho + \rho L^\dag_n L_n \right)\right]
\ee

In our case there are four Linblad operators $L_n$, corresponding to raising or lowering the fermion number at each end of the wire.
For example, the first and third terms of (\ref{eq:deltaH}) correspond to the Lindblad operator  $L_1 = c^\dag_R$, which
can be written in the 4-state basis
\be
|n_L, n_R \rangle =(  |0,0 \rangle, |1,1 \rangle |0,1 \rangle, |1,0 \rangle)^T
\ee
as
\ba \label{Lops1}
L_1 = c^\dag_R& =&
 \begin{pmatrix} 0 & 0 &  0 & 0 \\
0 &  0 & 0 & -1  \\
1 & 0 &0  &0 \\
0&0 &0 & 0 \\
\end{pmatrix} \ \ \ .
\ea

For $t_{23} =0$, the rate $\Gamma_1$ is given by
\be
\Gamma_1 =\frac{1}{\hbar}\, \left(   |\alpha_R^{(1)}|^2 + |\alpha_R^{(2)}|^2 \right )  \,\rho_R\, n_F(-2t_{34})  \ .
\ee
where $\rho_R$ is the density of states in the right reservoir.   Here $\alpha_R^{(1)}$ processes remove a particle of energy $- 2 t_{34}$ from the bath (probability $n_F(-2 t_{34})$) , while $\alpha_R^{(2)}$ processes create a particle of energy $2 t_{34}$ (probability $1 -n_F(2 t_{34}) = n_F(-2 t_{34})$).
(For the purposes of this discussion we are neglecting the effects of $t_{23}$, which is a reasonable approximation away from the avoided crossings, and is exact in the high-temperature limit.)  %Our numerical results, however, use an expression that is exact for any value of $t_{23}$, which is given in Appendix \ref{LindApp}).

Similarly, we have
\ba \label{Lops2}
L_2 = c_R& =&
 \begin{pmatrix} 0 & 0 &  1 & 0 \\
0 &  0 & 0 & 0  \\
0 & 0 &0  &0 \\
0& -1 &0 & 0 \\
\end{pmatrix}  \\
L_3 = c^\dag_L& =&
 \begin{pmatrix} 0 & 0 &  0 & 0 \\
0 &  0 & 1 & 0  \\
0 & 0 &0  &0 \\
1& 0 &0 & 0 \\
\end{pmatrix}  \ \ \ \ \
L_4 = c_L =
 \begin{pmatrix} 0 & 0 &  0 & 1 \\
0 &  0 & 0 & 0  \\
0 & 1 &0  &0 \\
0& 0 &0 & 0 \\
\end{pmatrix}  \nonumber
\ea
with the rates
\ba
\Gamma_2 &=&\frac{1}{\hbar}\,  \left(  |\overline{\alpha}_R^{(1)}|^2 + |\overline{\alpha}_R^{(2)}|^2\right )  \,\rho_R\, n_F(2t_{34}) \n
\Gamma_3 &=&\frac{1}{\hbar}\, \left(   |\alpha_L^{(1)}|^2 + |\alpha_L^{(2)}|^2 \right )  \,\rho_L\, n_F(-2t_{12}) \n
\Gamma_4 &=& \frac{1}{\hbar}\, \left(   |\overline{\alpha}_L^{(1)}|^2 + |\overline{\alpha}_L^{(2)}|^2 \right )  \,\rho_L\, n_F(2t_{12})
\ .
\ea
%Here we present results for the case $\Gamma_1/n_F(-2 t_{34})  = \Gamma_2 /n_F(2 t_{34})= \Gamma_3/n_F(-2 t_{12}) = \Gamma_4/n_F(2 t_{12})$.

The role of the temperature here is as follows.  If $T$ is small compared to the band gap energies $2 t_{12}$, $2 t_{34}$, then the fermion parity can change only by processes in which an excited state in, say, the parity even sector relaxes to a lower-energy state in the parity odd sector.  As we decrease $t_{12}$ from $t_{12}^{(0)}$ to $-t_{12}^{(0)}$ in the process described above, this means that parity-changing processes only occur for $t_{12}<0$, where in the absence of relaxation the system will be in an excited state.   If $T$ is large compared to the band gap energies, parity-changing processes can occur at any point along the sweep, enlarging the total probability that a relaxation event occurs during the sweep.
Because the high-temperature limit represents the worst-case scenario,  we will focus our attention on this limit, where it is natural to take $\Gamma_i \equiv \Gamma_1$, $\Gamma \equiv 4 \Gamma_1$.  (This is also quite possibly the limit pertinent to experiments, since although the experimental temperature is low, the distribution of quasi-particles in the bulk superconductor is typically not thermal\cite{AumentadoPoisoning,ShawPoisoning,FergusonPoisoning}).  The behavior of the system in the low-temperature limit is qualitatively similar, and is discussed in Appendix \ref{LowTApp}.

\begin{figure}
% \begin{tabular}{cc}
% \includegraphics[width=0.4\linewidth]{Fig1a.pdf} & \includegraphics[width=0.4\linewidth]{Fig1b.pdf} \\
 %\end{tabular}
% \begin{tabular}{llll}
% (a)& &   (b)& \\
% & \includegraphics[width=0.8\linewidth]{Overdamped.pdf}
%&  & \includegraphics[width=0.8\linewidth]{Overdamped2.pdf} \\
%  (c)& &  (d)& \\
%    & \includegraphics[width=0.8\linewidth]{Overdriven.pdf}
%&  & \includegraphics[width=0.8\linewidth]{SweetSpot.pdf} \\
\begin{tabular}{ll}
 (a)&  \\
 & \includegraphics[width=0.8\linewidth]{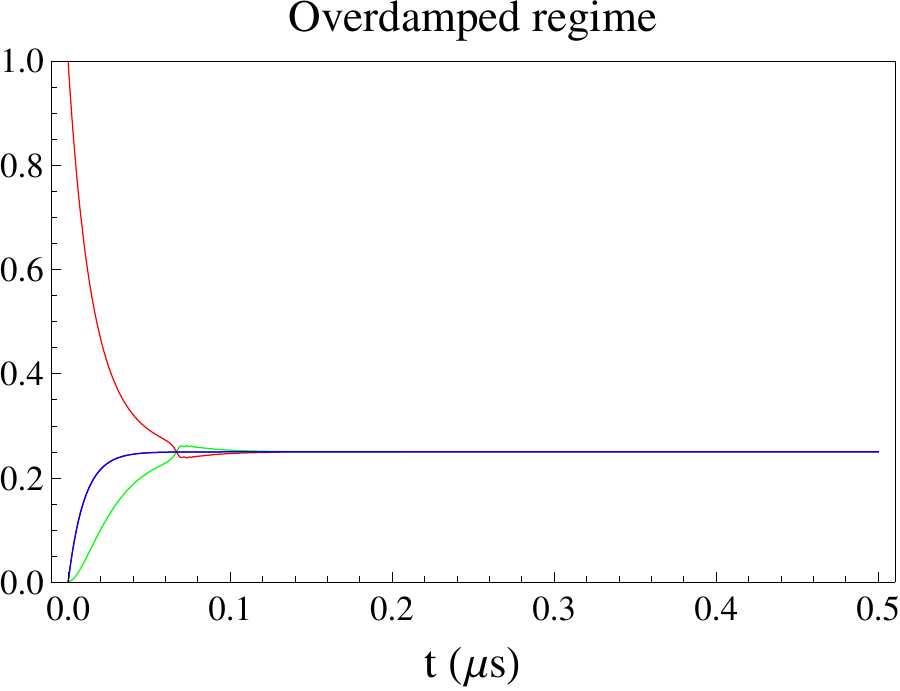}  \\
   (b)& \\
   & \includegraphics[width=0.8\linewidth]{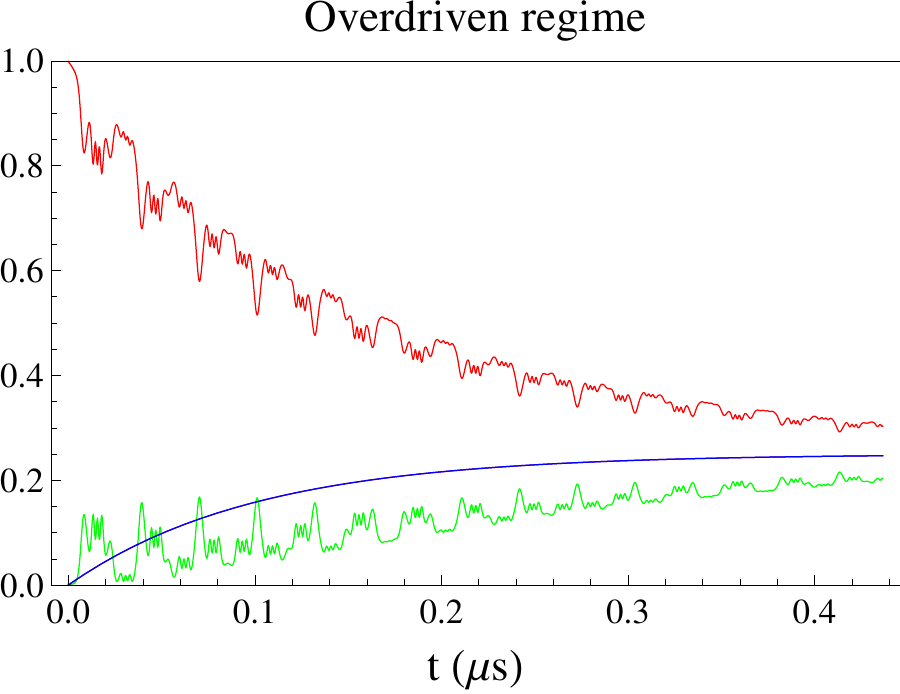} \\
  (c)& \\
  & \includegraphics[width=0.8\linewidth]{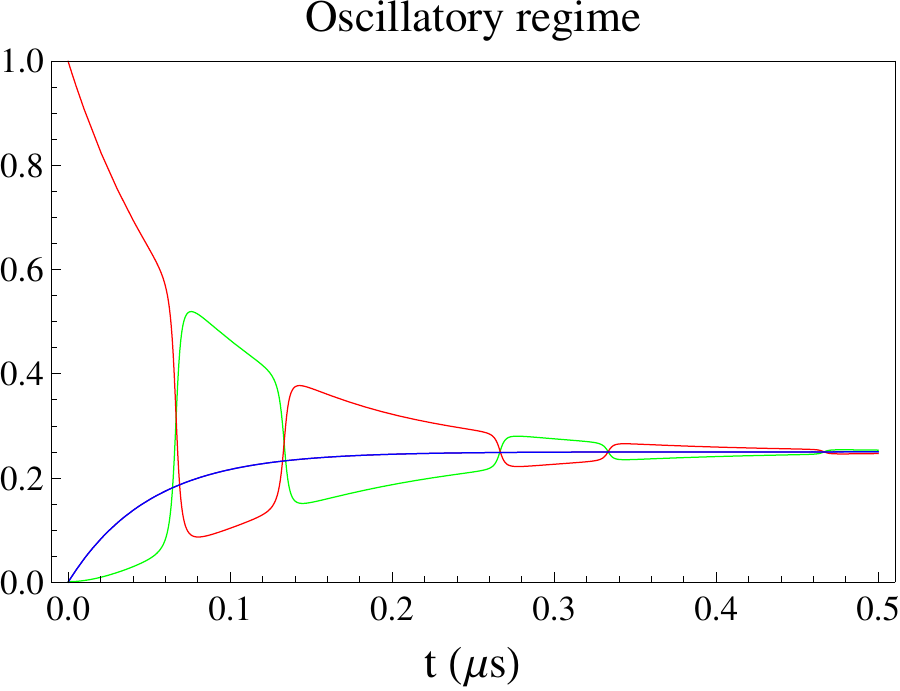} \\
   \end{tabular}
 \caption{Evolution of a density matrix initialized in the ground state for $t_{12} = t_{12}^{(0)}$, shown here over several periods of the oscillation $t_{12} =  t_{12}^{(0)} \cos \omega_0 t$.  For all plots we take $t_{23} = 0.1 t_{12}^{(0)} , t_{34} =0.5t_{12}^{(0)}$, and $\Gamma_i =  \Gamma/4$.  Here green is $\rho_{11}$, red is $\rho_{22}$, purple is $\rho_{33}$, and blue is $\rho_{44}$. %; these colors correspond to the color scheme for the corresponding states in Fig. \ref{Fig1}.   
 (a) For $\Gamma= 50 MHz, \omega_0/(2 \pi) =10 MHz, t_{12}^{(0)} /(2 \pi \hbar)=800 MHz $, the system is in the overdamped regime: the density matrix relaxes to its equilibrium value before oscillations can be observed.  (b) In the over-driven regime, shown here for $\Gamma= 5 MHz, \omega_0/(2 \pi) =64 MHz, t_{12}^{(0)} /(2 \pi \hbar)=160 MHz $, the oscillation undergoes many periods before the density matrix  thermalizes.  However Landau-Zener tunneling occurs even away from the avoided crossings, and
 there is no clear signature of Majorana-like correlations.  (c) In the oscillatory regime, (shown here for $\Gamma= 10 MHz, \omega_0/(2 \pi) =10 MHz, t_{12}^{(0)} /(2 \pi \hbar)=2.4GHz $), the probability of Landau-Zener tunneling at the avoided crossing is low, and the sweep is faster than the relaxation rate. % In this case the system switches reliably between predominantly $|1,1 \rangle$ and $|0,0 \rangle$ eigenstates twice per cycle, and several oscillations can be observed before the density matrix reaches its equilibrium value.   
 }
\label{RhoPlots}
 \end{figure}

The 16 components of the Master equation, together with an analytical discussion of their behavior in certain limiting cases, are presented in Appendix \ref{MasterApp}.  Given a choice of the temperature, hopping parameters, and rates, these equations can be solved numerically.  There are three time-scales in the problem: a fast time-scale $\tau_\epsilon$ set by the mean value of the energy gaps shown in Fig. \ref{Fig1}, the period $\tau_\omega = 1/\omega_0$ of oscillation of $t_{12}$, and the parity time $\Gamma^{-1}$.  For or our purposes, the solutions to (\ref{MasterEq}) fall into three regimes, depending on the
relative sizes of these three time-scales.  If $\Gamma^{-1}$ is fast compared to $\tau_\omega$, then a system that begins in the eigenstate $|1,1 \rangle$ at $t_{12} = t_{12}^{(0)}$ will relax to an equilibrium distribution
before the avoided level crossing, and no oscillations in $n_R$ can be observed.  We will call this the overdamped regime (Fig. \ref{RhoPlots}a).  If $\tau_\omega$ is fast compared to $\tau_\epsilon$ (Fig. \ref{RhoPlots}b), Landau-Zener tunneling occurs even away from the avoided crossing.  In this case the dynamics can be quite complex, but we do not reliably find oscillations with a clear signature of Majorana correlations.   We call this the over-driven regime.
   In between these two regimes, if $\Gamma^{-1} \gg \tau_\omega \gg 1/(2  \hbar \ t_{23})$, the system passes quasi-adiabatically through the avoided level crossing, and one or more oscillations occur in the density matrix before equilibrium is reached.  %probability of finding the system in a state with $n_R =0$ before it relaxes to equilibrium.
 We call this the oscillatory regime (Fig. \ref{RhoPlots}c); it is here that the parity experiment can detect Majorana-like long ranged entanglement.

 Assuming that the driving frequency $\omega_0$ can be easily adjusted to produce optimal results, the success of the parity experiment in a given wire system will depend on whether the time-scale  $t_{ij}/\hbar$ of the Majorana interactions is fast or slow relative to the decay rates $\Gamma_i$.    Estimates\cite{DasSarmaSmoking,RainisPRBB87} based on parameters relevant to the experiments of Ref. \onlinecite{MourikScience336} suggest  that values of $t_{12},t_{34}$ up to $30\  \mu eV$ are not unrealistic in short ($\sim 1 \ \mu m$) wires.   This gives $t_{ij} /(2 \pi \hbar) \sim \mathcal{O}( 2.5 GHz)$.  Conversely relaxation rates are expected to be, at worst, $\Gamma \sim \mathcal{O}( 10 MHz)$\cite{RainisLossPRB85.174533}.  For parameter values in this rage, the system is in the oscillatory regime for $\omega_0/(2 \pi) =10 MHz$ (Fig. \ref{RhoPlots} c).
%Even if the couplings $t_{ij}/\hbar \sim \mathcal{O}( 1 GHz)$ are significantly smaller than anticipated, driving the system at higher frequencies can still produce a positive result
Of course the precise values of $t_{ij}$ will depend on the lengths of the respective wire segments, as well as the effective wire Hamiltonian in the superconducting regime.  The important point is that
the parity experiment can be performed in the oscillatory regime provided that $\Gamma$ can be made small\footnote{In fact, it is not imperative that $\omega_0$ be small compared to the splitting $2 t_{23}/\hbar$ at the avoided crossing, though in this case the dynamics is more complicated and several cycles of $t_{12}$ are required to observe switching.} relative to $t_{23}/\hbar$ %; otherwise the system is necessarily either in the over-driven or over-damped regime.
 -- a prospect that seems quite realistic, based on current theory\cite{RainisPRBB87,DasSarmaSmoking,RainisLossPRB85.174533}

% If the sweep is fast compared to $\hbar/(2 t_{23})$, but slow compared to $\tau_\epsilon$, %it is also possible in some cases to observe Majorana correlations.  In this regime
% the system has a high probability of Landau-Zener tunneling at the avoided crossings, and hence remaining in the $|1,1 \rangle$  state rather than relaxing to the $|0,0 \rangle$ ground state.  However switching is observed after $n$ cycles of $t_{12}$ %are required to interchange the relative magnitudes of $\rho_{11}$ and $\rho_{00}$,
%where $n$ increases with  the tunneling probability.  Thus Majorana correlations can be observed if $\tau_d$ is long relative to $n \tau_{\omega}$.

\section{Measuring the parity response} \label{ParityMeas}

We now consider what signal this parity response will give in the experimental system.
First, in order to develop intuition, we attempt a simple (``poor man's")
quantum jump \cite{Plenio98QuantumJumps} analysis of the system dynamics, shown 
in Fig.~\ref{fg:ParityTimeFig}.  These curves are obtained using the following algorithm.
Assuming that the fermion parity is strictly conserved (i.e. $\Gamma=0$), the expectation value $\langle \hat n_R\rangle$ is calculated  
for each of the four possible initial states at $t=\tau_0$, using the evolution (Eq. \ref{MasterEq}) of the density matrix.  Each curve shows periodic
oscillations in which $\langle \hat n_R\rangle$ changes smoothly between (approximately) 0 and 1.
If there were no parity relaxation, only one of the curves (determined by the initial state) would be observed.
Parity relaxation processes induce (Poissonian random) jumps between the 4 curves.
The higher the rates $\Gamma_n$, the more frequent the jumps.
In Fig.~~\ref{fg:ParityTimeFig} we show two examples, one of rare (Fig.~~\ref{fg:ParityTimeFig}a) and one of frequent (Fig.~~\ref{fg:ParityTimeFig}b) jumps.
Such curves would roughly correspond to a typical measurement trace of $\hat n_R$,
provided the measurement is not strong enough to distinguish between
$n_R=0$ and $n_R=1$ on the time-scale required to  adiabatically traverse the avoided crossings.

\begin{figure}
% \begin{tabular}{cc}
% \includegraphics[width=0.4\linewidth]{Fig1a.pdf} & \includegraphics[width=0.4\linewidth]{Fig1b.pdf} \\
 %\end{tabular}
 \begin{tabular}{ll}
 (a)& \\
 & \includegraphics[width=0.8\linewidth]{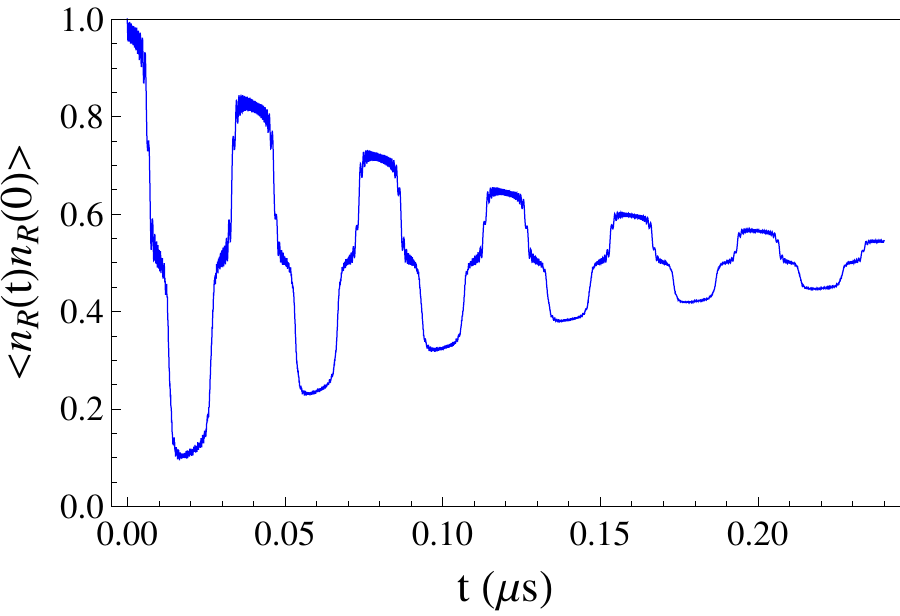} \\
  (b)& \\
  & \includegraphics[width=0.8\linewidth]{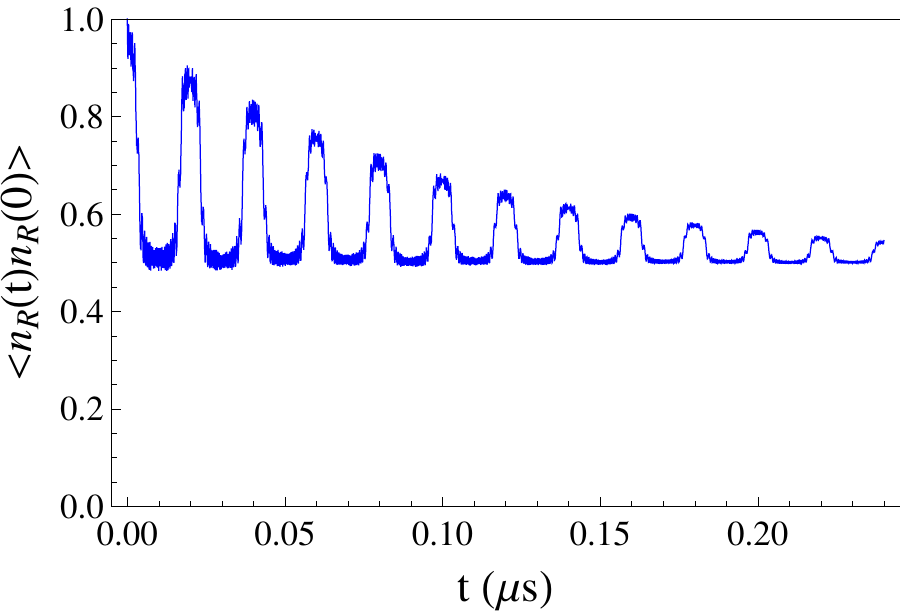} \\
%  (c)& \\
%  & \includegraphics[width=0.8\linewidth]{NCorPlots3.pdf}
   \end{tabular}
 \caption{ The autocorrelation function $\langle n_R (\tau) n_R(\tau_0) \rangle$ when $t_{12}(\tau_0) =t_{12}^{(0)}$ (a) and when  $t_{12}(\tau_0) =0$ (b).   Both plots are shown for $t_{12}^{(0)}/(2 \pi \hbar) = 2.4 GHz, t_{23}/(2 \pi \hbar) =0.36 GHz, \omega_0/(2 \pi) = 25 MHz, t_{34}/(2 \pi \hbar) = 1.2 GHz$, $\Gamma =0.01GHz$.   }
\label{MeasCorFig}
 \end{figure}

A more detailed analysis can be made by examining the autocorrelation function 
\be
\langle n_R (\tau) n_R(\tau_0) \rangle - \langle n_R (\tau_0)\rangle^2 \ \ \ .
\ee  
In this case we will assume that the measurement couples strongly to the system, such that measuring $n_R$ collapses the wave-function.  Between measurements the coupling to the measurement device is switched off and the system evolves according to the dynamics described above.
A strong measurement can be useful since in practice the system will initially be described by its equilibrium density matrix, which at high temperature is non-dynamical even as $t_{12}$ is varied.  Even at  temperatures that are high relative to the $t_{ij}$, however, the Majorana correlations can be seen by  %$n_R$ both at $t=n /\omega$, where  $t_{12} = t_{12}^{(0)}$, and
%performing two subsequent measurements of $n_R$ separated by a time interval $\tau$ as $t_{12}$ is varied.
measuring the autocorrelation function.  

The autocorrelation function can be extracted from the time-evolution of the density matrix as follows.  In the markovian limit, using the
quantum regression theorem \cite{Lax63,GardinerZollerBook} we obtain
\ba \label{AutoCorr}
\langle n_R (\tau) n_R(\tau_0) \rangle = {\rm Tr} \left[\hat n_R\, U(\tau,\tau_0) \,\hat n_R \,\rho(\tau_0) \right]
\ea
Here $U(\tau,\tau_0)$ is the dissipative evolution (super) operator obtained by integrating the master equation
Eq.~\ref{MasterEq}. Indeed,  Eq.~\ref{MasterEq} can be rewritten by ``vectorizing" the density matrix $\rho$, i.e.,
by writing its rows one after another into a vector of length 16. This gives $\dot \rho = {\cal L}\rho$, where
the Lindblad (super) matrix ${\cal L}$ is given by
\ba
\label{MasterEqVec}
{\cal L} &=&- \frac{i}{\hbar} (H\otimes \hat 1 - \hat 1 \otimes H^*)
\nonumber\\
&+& \sum_n \,\Gamma_n\,\left[ L_n \otimes L^{*}_n - \frac{1}{2} \left( L^\dag_n L_n \otimes \hat 1 + \hat 1\otimes  L^T_n L^*_n \right)\right]\ .
\nonumber\\
\ea
The evolution operator is then given by
\be
U(\tau,\tau_0) = T \exp\left[\int\limits_{\tau_0}^{\tau} {\cal L(\tau')} \,d\tau'\right]\ .
\ee

In other words, we calculate the autocorrelation function (\ref{AutoCorr}) by first projecting the equilibrium density matrix onto a density matrix with a definite value of $n_R = 1$ at time $\tau_0$.  Next we evolve the resulting matrix $\hat n_R \rho(\tau_0)$ in time using the master equation (\ref{MasterEq}).
To obtain the autocorrelation function, we evaluate the probability of measuring $n_R =1$ at the end of this time evolution.  
In the presence of a driving force the Lindblad matrix ${\cal L}$ is time dependent, such that the autocorrelation function depends not only on the elapsed time $\tau-\tau_0$, but also on the relative phase $\phi$ between the oscillation of $t_{12}$ and the time $\tau_0$ at which the evolution is initiated.

The expected autocorrelation measurement for the parity experiment is shown in Fig. \ref{MeasCorFig}.
To understand this plot, consider first the situation with $\Gamma = 0$, and in the oscillatory regime, with $\phi=0$ (meaning that the autocorrelation function is phase-locked to the maximum of $t_{12} = t_{12}^{(0)}$; Fig. \ref{MeasCorFig}a).  Measuring $n_R =1$ projects the system onto a mixture of the two
states $|1,1 \rangle$ and $|0,1 \rangle$.   If $\tau$ is less than approximately $1/8$ of the period, the system has not passed through either of the avoided crossings, and a second measurement detects $n_R=1$
again with a high probability.  For $1/ (8 \omega_0) <  \tau < 3/(8 \omega_0)$, we have passed through the avoided crossing in the parity-odd sector, but not that of the parity even sector.
 Hence in the oscillatory regime, the state $|0,1 \rangle$ has predominantly switched to the state $|1,0 \rangle$, while the state $|1,1 \rangle$
has not switched.  In this case the autocorrelation is very small, as the probabilities of measuring $n_R =0,1$ are essentially equal.  For $3/(8 \omega_0) < \tau < 5/8 \omega_0$, we have passed through both avoided
crossings, and our mixture of states $|1,1 \rangle, |0,1 \rangle$ has been replaced by a mixture of the states $|0,0 \rangle$ and $|1,0 \rangle$.   At these times, therefore, the autocorrelation is negative.  Hence the autocorrelation function oscillates at the frequency $\omega_0$.

If $\phi = \pi/2$ (i.e. the autocorrelation function is out of phase with $t_{12}$, and $t_{12}(\tau_0) =0$, as in Fig. \ref{MeasCorFig}b), then the even parity sector switches from $|1,1 \rangle$ to $|0,0 \rangle$ {\it and back} before any switching occurs in the parity odd sector.  In this case the autocorrelation function is never negative, and oscillates at a frequency of approximately $2 \omega_0$.  
For finite relaxation rates the autocorrelation function  (\ref{AutoCorr}) oscillates as described above, but the oscillation is contained in an exponentially decaying envelope with time constant $\Gamma^{-1}$.

\begin{figure}
 \begin{tabular}{cc}
 (a) & \\
  &\includegraphics[width=0.8\linewidth]{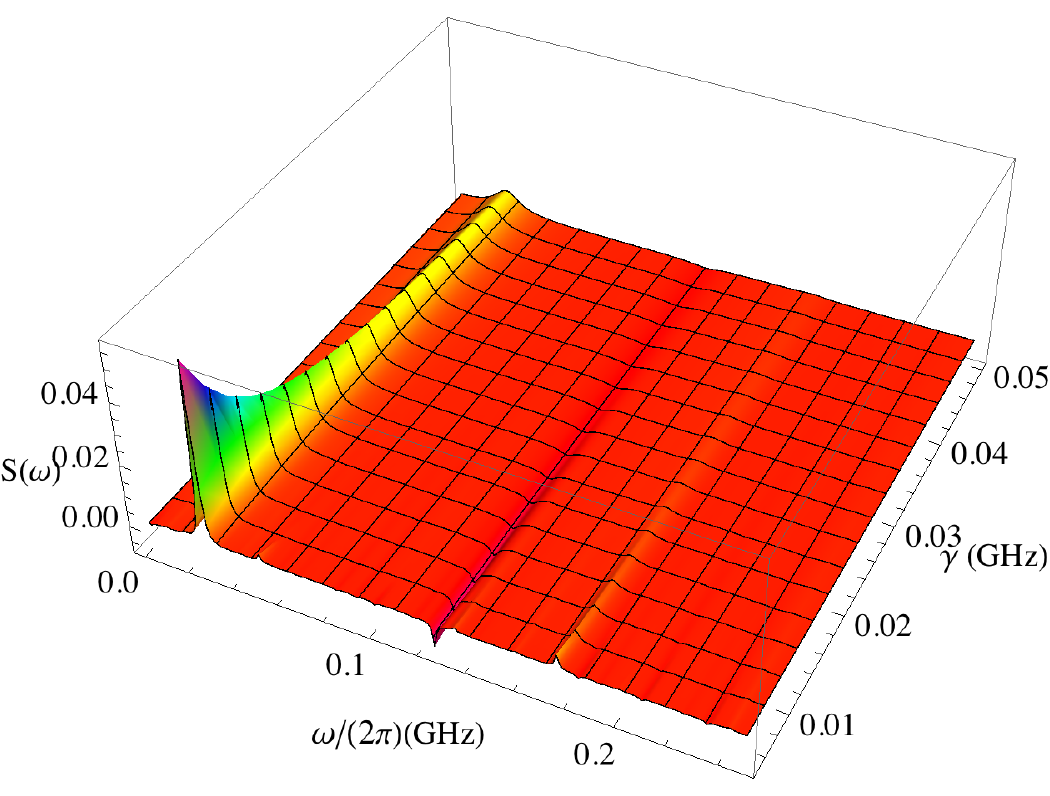} \\
  (b) & \\
  &  \includegraphics[width=0.8\linewidth]{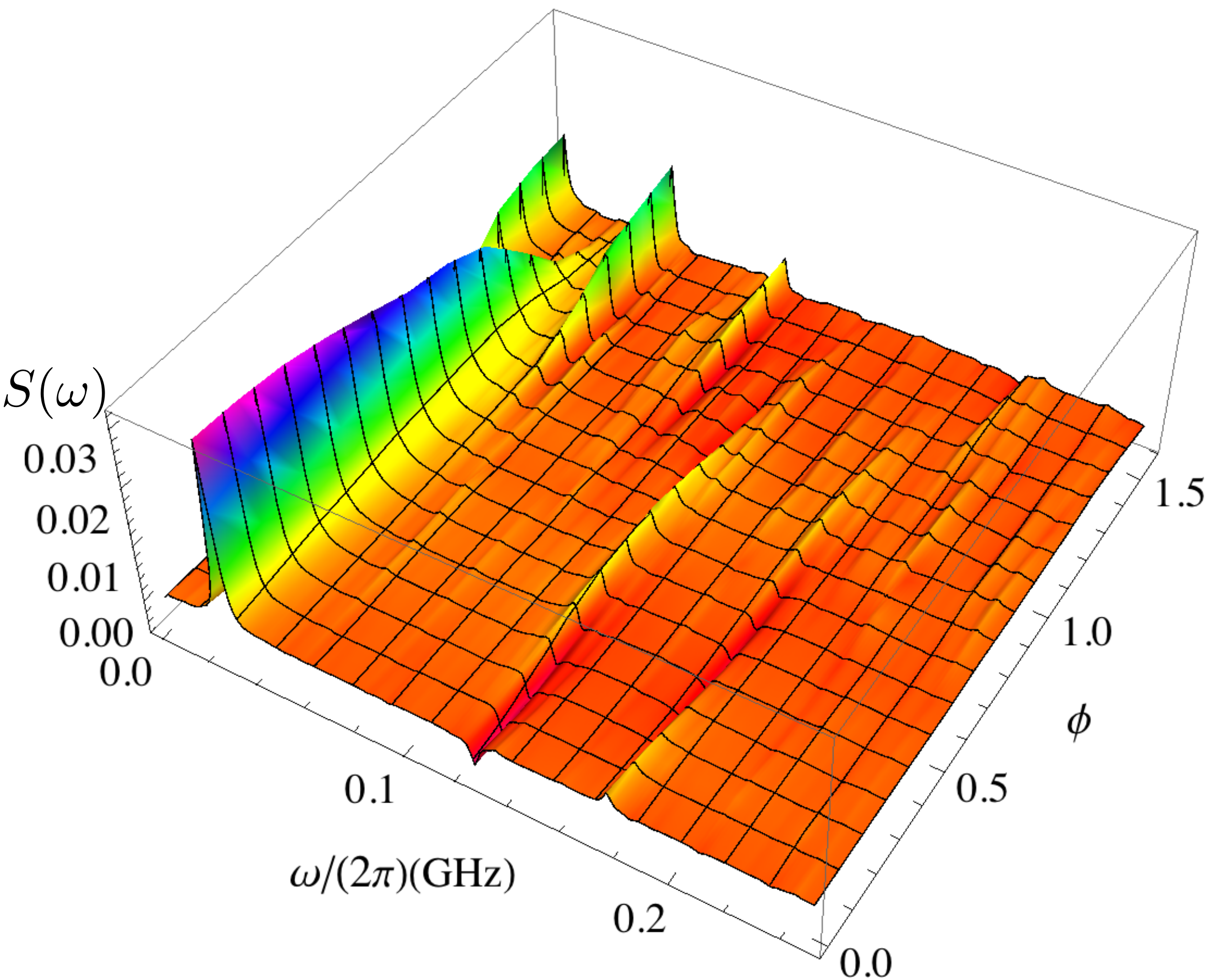} \\
  \end{tabular}
 \caption{The real part ($S(\omega)$)
  of the Fourier transform of $\langle n_R(\tau) n_R(0) \rangle$ as a function of (a) the relaxation rate $\Gamma$, for $\phi=0$, and (b) the phase offset $\phi$ between the maximum value of $t_{12}$ and the autocorrelation function, for $\Gamma = 100 MHz$.  Both plots are shown for $t_{12}^{(0)}/(2 \pi \hbar) = 2.4 GHz, t_{34}/(2 \pi \hbar)  = 1.2 GHz, t_{23}/(2 \pi \hbar)  = 0.36 GHz, \omega_0/(2 \pi) = 25 MHz$.   }
\label{FourierFig}
 \end{figure}

Fig. \ref{FourierFig} shows the Fourier transform of the autocorrelation function.
The height of the Fourier peak decreases as the relaxation rate $\Gamma$ is increased.  Experimentally, we expect $\Gamma < \mathcal{O} (10 MHz)$, where the Fourier peak is relatively pronounced, as can be seen for $\phi =0$ in Fig. \ref{FourierFig} (a). As the phase offset $\phi$ between the maximum value of $t_{12}$ and the autocorrelation measurement is varied, both the height and position of the Fourier peaks varies.  At $\phi =0$ we see a large peak centered approximately at the driving frequency $\omega_0$, and a second smaller peak at high frequencies (due to the fact that eigenstates of $n_R$ are not exact eigenstates of the Hamiltonian).  As $\phi$ approaches $\pi/2$, the height of the peak at $\omega_0$ decreases, and new peaks at a frequency of approximately $2 \omega_0$ appear (Fig. \ref{FourierFig} (b) ).

\section{Measurements at constant $t_{12}$: noise spectrum and correlations} \label{NoiseSec}

The parity experiment that we have discussed thus far requires driving oscillations of the coupling $t_{12}$ between the two Majoranas on the left wire segment.  
Here we turn our attention to what can be learned by measuring the autocorrelation function 
%in the set-up shown in Fig. \ref{WireFig} while holding $t_{12}$  constant in time. 
 at a fixed value of $t_{12}$. For constant $t_{12}$, the autocorrelation function simply probes the rate at which $n_R$ relaxes to its equilibrium value.  In this section, we will explore how these relaxation processes can be used both to measure the parity time $\Gamma^{-1}$, and to probe the correlation between the two wire segments.
 
It is worth emphasizing that measuring the  parity time is of practical value in and of itself.  $\Gamma^{-1}$ is expected to be quite long in an ideal isolated system, since in theory at low temperature in the superconductor that is in proximity to the wire there are only pairs of electrons. In practise, however, this time is affected by various experimental factors, and may depend on many details of the system in question, such as tunneling from a metallic lead with free electrons, %the existence of free quasi particles 
quasiparticle poisoning in the superconductor, % at finite temperature, 
or the presence of %impurity states and magnetic 
impurities. (For a recent review see Ref.~\onlinecite{Alicea2012RPP} and references therein.)  Hence  the parity relaxation rate in a particular device is difficult to predict, and in practise must be measured.

We begin with a slightly simpler experimental set-up than that shown in Fig. \ref{WireFig}, in which we have only one wire segment.  (In our original set-up, this amounts to taking $t_{23} =0$, in which case the right- and left- wire segments are completely decoupled.)  
If the coupling $t_{34}$ between the two Majoranas in our wire segment is held constant in time, then the autocorrelation function in this wire is:
\be \label{AutoCor1Wire}
\langle n (\tau) n(\tau_0) \rangle  -  \langle n(\tau_0) \rangle^2 = e^{- \Gamma |\tau-\tau_0|}  \frac{e^{ -2 \beta t_{34} }}{ (1 + e^{ -2 \beta t_{34} } )^2} 
\ee
with $ \langle n(\tau_0) \rangle = 1/(1 + e^{- 2 \beta t_{34} })$.  
%{\bf AS: Fiona, probably the time dependence should be $e^{- \Gamma (\tau-\tau_0)} $?}
Fourier transforming this gives the noise spectrum, whose real part is given by:
\be
S(\omega) =\frac{e^{ -2 \beta t_{34} }}{ (1 + e^{ -2 \beta t_{34} } )^2}   \frac{ 2 \Gamma}{\Gamma^2 + \omega^2}
\ee
%{\bf AS: Fiona, I suspect its $2\Gamma / (\Gamma^2 + \omega^2)$}
In the high-temperature limit $\beta t_{34}  \ll 1$, the pre-factor is approximately $1/4$ and one can determine $\Gamma$ by measuring the height of the zero-frequency peak, via
\be
\Gamma = 2 \left(  \lim_{\omega \rightarrow 0} S(\omega) \right)^{-1}
\ee
(In practise, the measurement need not be carried out at zero frequency, but merely at frequencies that are small compared to the parity time $\Gamma$).  

Next, consider a setup with coupled left and right wire segments.  Unlike the situation described above, the operator $n_R(\tau_0)$ does not project the system onto an eigenstate, and the resultant dynamics is more complicated. 
For $t_{12}$ constant, in the high-temperature limit the autocorrelation function has a relatively simple analytic form:
\ba \label{AutoCorr2}
\langle n (\tau) n(0) \rangle  -  \langle n(0) \rangle^2 &=& \frac{e^{ - \Gamma |\tau |} }{4} \left [  1 - 2t_{23}^2 \left ( \frac{1}{\epsilon_+^2 } + \frac{1}{\epsilon_-^2} \right ) \right ] \n 
&&  +  t_{23}^2 \frac{e^{ - \Gamma |\tau| } }{2}  \left ( \frac{\cos ( \frac{\epsilon_+}{\hbar}  \tau)  }{\epsilon_+^2 } + \frac{\cos (\frac{\epsilon_-}{\hbar}   \tau) }{\epsilon_-^2} \right ) \nonumber 
\ea
where $\epsilon_\pm = 2 \sqrt{ (t_{12}\pm t_{34})^2+ t_{23}^2}$, and we have assumed $\Gamma_L = \Gamma_R$.  Hence the noise spectrum differs from that of a single wire segment in two ways.  First, the height of the low-frequency peak decreases by an amount of order $2 t_{23}^2/ \text{min}(\epsilon_+, \epsilon_-) $.  Second,  finite frequency peaks of the same order appear, at frequencies $\epsilon_{\pm}/ \hbar$. 

\begin{figure}
 %\begin{tabular}{cc}
 %(a) & \\
  %&
  \includegraphics[width=0.8\linewidth]{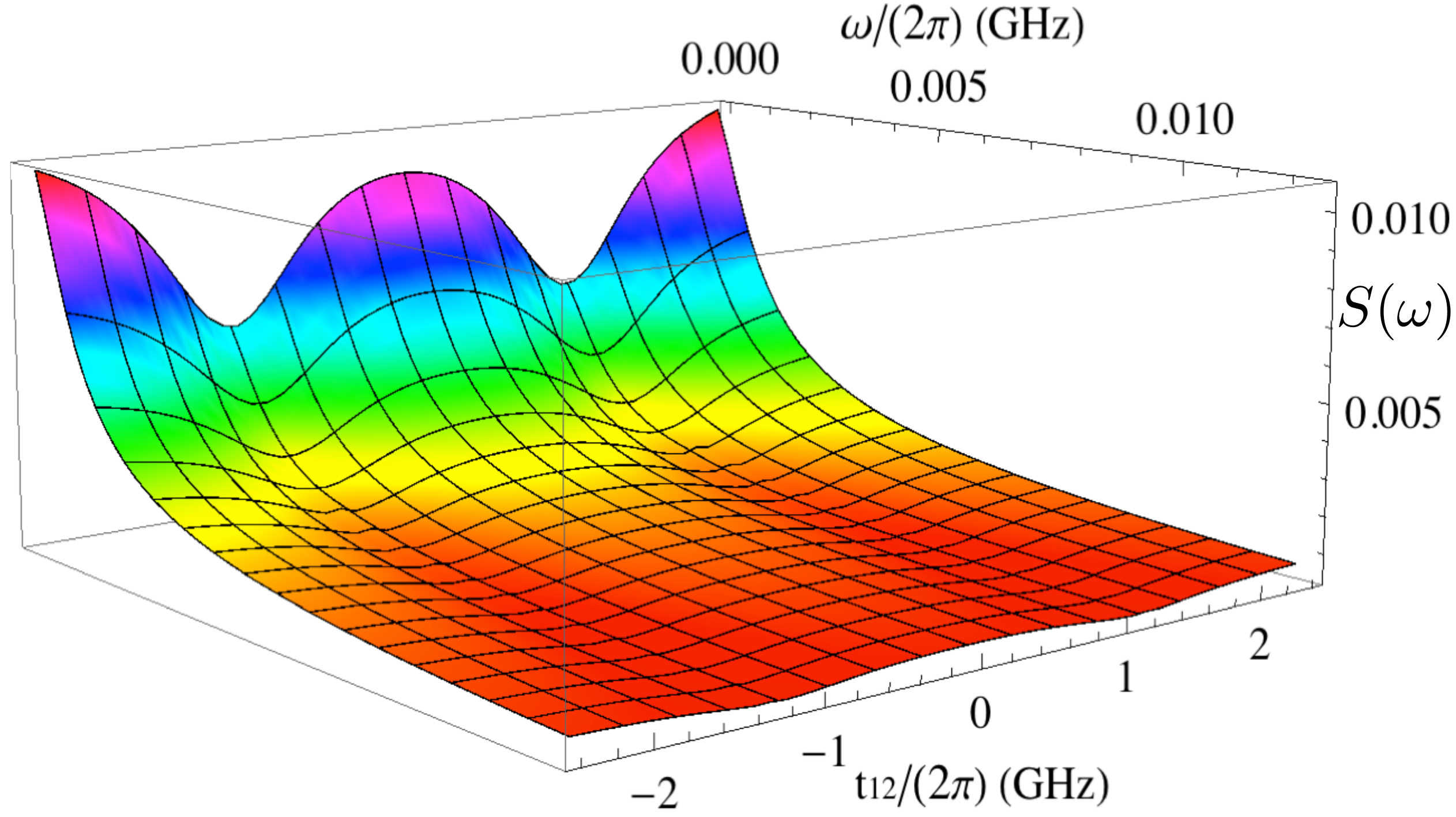} %\\
%  (b) & \\
%  &  \includegraphics[width=0.8\linewidth]{} \\
%  \end{tabular}
 \caption{The noise spectrum as a function of $t_{12}$ in the high-temperature limit, shown for $t_{34}/(2 \pi \hbar)  = 1.2 GHz, t_{23}/(2 \pi \hbar)  = 0.36 GHz, \Gamma = 250 MHz$.   The peaks in the low-frequency noise spectrum show marked dips near the avoided crossings, where finite-frequency noise appears.   %(b) In the low-temperature limit the low-frequency noise spectrum shows a peak near $t_{!2} =0$.  
 }
\label{NoiseFig}
 \end{figure}

The upshot is that the noise spectrum (Fig. \ref{NoiseFig}) shows signatures of correlations between the Majoranas in the left- and right- wire segments. 
If $| t_{23}| \ll \epsilon_{+,-}$, then in practise the noise spectrum differs only slightly from that of the isolated wire segment described above.  However, near the avoided crossings $t_{12} =  \pm t_{34}$, the noise spectrum is distinctly different from the uncoupled wire case: the height of the zero-frequency peak will decrease from  approximately $1/ (4 \Gamma)$ to approximately $1/ ( 8\Gamma)$, and a finite frequency peak of approximately height approximately $1 / (16 \Gamma)$ appears.  
 We believe that this reduction by a factor of $2$ in the low-frequency noise $S(\omega)$ near the avoided crossings should be experimentally observable, and give an enticing hint that Majorana correlations exist in the nanowire systems.

These variations in the noise spectrum near the avoided crossings have an intuitive physical origin.  Away from the avoided crossings the states $| 1 ,1 \rangle, |0,1 \rangle$ that $n_R(\tau_0)$ projects onto are very close to being eigenstates of the system, and the autocorrelation function (\ref{AutoCorr2}) is 
dominated by parity relaxation processes in the right-hand wire segment.  Near the avoided crossing at $t_{12} = t_{34}$, however (see Fig. \ref{Fig1}), the eigenstates in the parity-odd sector are
$ |\psi_{\pm} \rangle =\frac{1}{\sqrt{2} } \left( |0,1\rangle \pm |1,0\rangle \right)$.  
%{\bf AS: Fiona, I propose to write this differently. New text:}
Thus, on time scales longer than $t_{34}^{-1}$ the observable $n_R$ takes the value of $1/2$ in these two states. 
At high temperature the system spends half of the time in the states $ |\psi_{\pm} \rangle =\frac{1}{\sqrt{2} } \left( |0,1\rangle \pm |1,0\rangle \right)$ with $n_R = 1/2$, 
one quarter of the time in the state  $|1,1 \rangle$ with $n_R = 1$, and one quarter of the time in the state $|0,0 \rangle$ with $n_R=0$. Thus for the 
same-time fluctuations we obtain
\begin{equation}
\int_{low freq.} \frac{d\omega}{2\pi}\,S(\omega)= \langle n_R^2\rangle - \langle n_R\rangle^2 = \frac{3}{8}-\frac{1}{4}= \frac{1}{8}\ .
\end{equation}
This is in contrast to the situation away from the avoided crossing point where the system spends half of the time in state $n_R =1$ 
and half of the time in state $n_R =0$. Thus $\langle n_R^2\rangle - \langle n_R\rangle^2=1/4$. Clearly, half of the spectral weight 
shifts to the higher frequencies (or order $t_{23}$) at the avoided crossing points. 
%{\bf End of new text}
%
%
%******
%
%{\bf Old text}
%Thus a system initially in the state $|0,1 \rangle$ evolves in time according to
%\be
%|\psi (\tau) \rangle = \frac{1}{\sqrt{2}} \left( |\psi_+ \rangle e^{ i 2 t_{23} \tau } +  |\psi_- \rangle e^{ -i 2 t_{23} \tau } \right )
%\ee
%so that 
%\be
%\langle \psi (\tau)| n_R (\tau)|\psi (\tau)  \rangle = \frac{1}{4} \left( 1 + \cos (4  t_{23} \tau )  \right )
%\ee
%Hence if there were no relaxation processes, after averaging over the oscillations, we would find:
%\be
%\langle n_R(\tau) n_R (\tau_0) \rangle \approx 
%\frac{1}{4}  \langle 1,1 | n_R |1,1 \rangle + \frac{1}{2} \langle \psi (\tau)| n_R (\tau)|\psi (\tau)  \rangle = \frac{3}{8}
%\ee
%as opposed to the value of $1/2$ obtained away from the avoided crossings.  (In the high-temperature regime $\langle n_R (\tau_0) \rangle^2 = 1/4$, irrespective of $t_{12}$).  This is the origin of the dip in the height of the zero-frequency peak near these avoided crossings.
%{\bf End of old text}
%
%***

We note that near the avoided crossings, relaxation processes in both the left and right wire segments contribute equally to the total relaxation rate -- which is not the case for more generic values of $t_{12}$ where relaxation is dominated by processes occurring in the right-hand wire segment only.  Consequently if the relaxation rate in the left wire segment is significantly larger than that in the right wire segment, the effective relaxation rate at the avoided crossings will be considerably higher than for other values of $t_{12}$, making the decrease in $\lim_{\omega \rightarrow 0} S(\omega)$ even more pronounced.

In the low-temperature regime the height of the zero-frequency peak in the noise is significantly diminished, as is apparent in Eq. (\ref{AutoCor1Wire}).  This is because the system will relax into its ground state, which is essentially an eigenstate of $n_R$ for most values of $t_{12}$.   In this limit it is difficult to extract any useful information about the system from the low-frequency behavior of the noise spectrum.

\section{Summary}

We have presented an experimental protocol that can be used to detect long-ranged entanglement between pairs of Majorana fermions at the ends of 1D superconducting wires without actually carrying out experimentally challenging brading operations.  Our suggested experiment requires a single wire (of length $\sim 1-2 \ \mu m$) with two regions of topological superconductor separated by a normal region, and the capacity to measure the fermion parity in one of the two topological wire segments.  Even for conservative estimates of the parity relaxation time ($\Gamma \approx 10 MHz$), realistic parameter estimates\cite{DasSarmaSmoking,RainisPRBB87} suggest that this set-up can be used to detect correlations between the fermion parity in the left- and right- hand wire segments which would be strongly suggestive of the sought- after Majorana zero modes in these systems.

{\bf Acknowledgements} We would like to thank A. Haim for enlightening discussions.  FJB and YO are grateful to KITP (NSF Grant no. PHY11-25915) for its hospitality.
AS is grateful to the Weizmann Institute of Schience (Weston Visiting Professorship) and the
BMBF Project RUS 10/053 ``Topologische Materialien f\"ur Nanoelektronik''. YO is grateful for support by  DFG, TAMU  and ISF grants.

\appendix

\section{Dynamics at low temperature} \label{LowTApp}

Here we outline the expected outcome of the parity experiment in the ``low-temperature" regime.  We note that the effective temperature is that of the bath, and not of the wire itself.  
The main interesting feature of the low-temperature regime is that at bath temperatures low relative to $t_{34}$, it is possible to observe Majorana correlations at signifincantly higher values of $\Gamma$ than at high temperatures.
In fact, even for values of $\Gamma$ that are somewhat larger than $\omega_0$, is possible for the autocorrelation function to display persistent oscillations.

\begin{figure}
\begin{tabular}{ll}
a & \\
&   \includegraphics[width=0.8\linewidth]{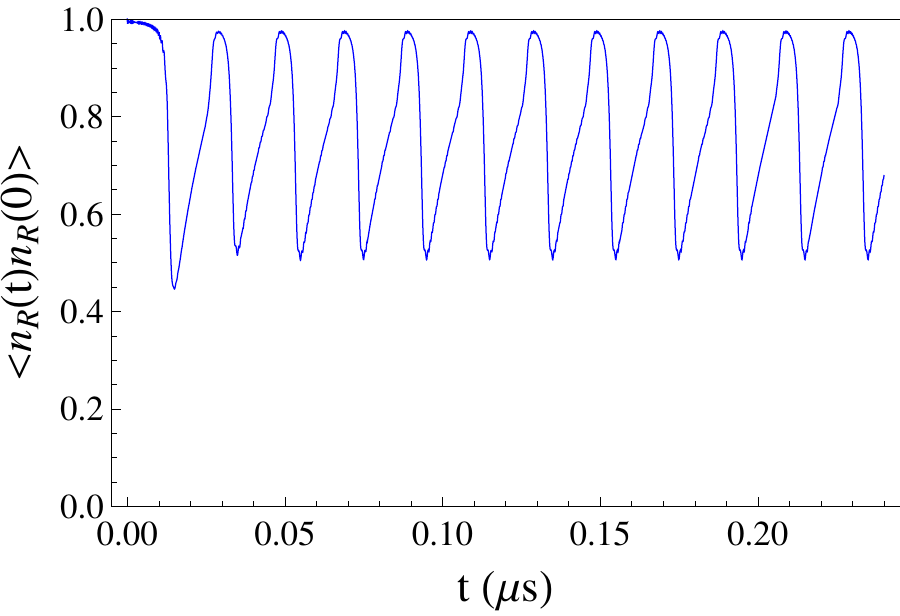} \\
b & \\
&   \includegraphics[width=0.8\linewidth]{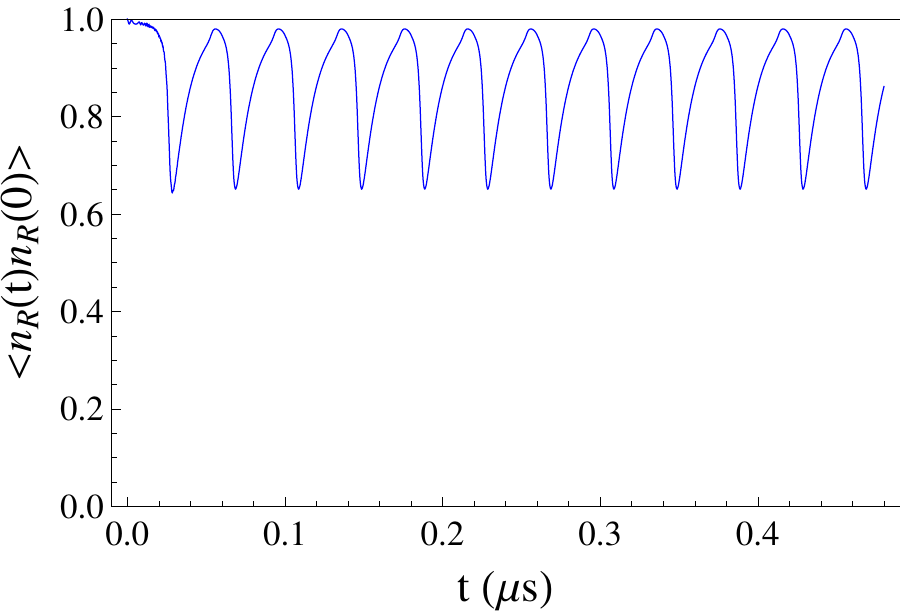} \\
\end{tabular}
 \caption{The autocorrelation function in the low temperature limit, shown here for $ t_{12}^{(0)}/(2 \pi \hbar) =2.4 GHz, t_{34}/(2 \pi \hbar)  = 1.2 GHz, t_{23}/(2 \pi \hbar)  = 0.36 GHz$, and (a) $\Gamma = .1 GHz, \omega_0/(2 \pi) = 25 MHz$; (b)  $\Gamma = .1 GHz, \omega_0/(2 \pi)= 12 MHz$.  Both plots are shown for $\phi=0$. }
\label{TDensityPlot}
 \end{figure}

Consider initializing the system in its ground state $|1,1 \rangle$ at $t_{12} = t_{12}^{(0)}$.  This remains the ground state until $t_{12}$ becomes negative, so the decay from the parity-even to parity-odd sector begins at time $t = 1/4 (2 \pi /\omega_0)$.  If relaxation is not complete by the time the avoided crossing is reached an eigth of a cycle later, there is a significant probability of finding the system in the first excited state, and consequently measuring $n_R =0$.  If the decay rate is relatively fast, then by the time $t_{12}$ again becomes positive, $3/4$ of the way through the cycle, the system has relaxed to its ground state for negative $t_{12}$, which is $|0,1 \rangle$, and $\langle n_R \rangle \approx 1$.   As $t_{12}$ increases the process repeats itself: provided that the system has not totally relaxed to the ground state by the time $t_{12} = t_{34}$, we will again briefly have a significant chance of finding $n_R =0$.  Because the system always relaxes to its ground state, this cycle can repeat itself indefinitely, and the autocorrelation function displays persistent oscillations, as shown in Fig. \ref{TDensityPlot}.

The crucial point here is that the relaxation rate need only be slow enough that the system does not relax over the course of $1/8$ of a cycle, rather than over a minimum of $1/2$ a cycle for the high temperature limit.  Since the oscillations are persistent, the Fourier peak is also sharper.

\section{Master Equation} \label{MasterApp}

Here we present the details of the time evolution of the density matrix dictated by Eq. (\ref{MasterEq}).  
\begin{widetext}
We may parametrize the density matrix as follows:
\be
\rho(t) = \begin{pmatrix} \rho_{11}(t) & \rho_{12}(t) & \rho_{13}(t) & \rho_{14}(t) \\
\rho^*_{12}(t) & \rho_{22}(t) & \rho_{23}(t) & \rho_{24}(t) \\
\rho^*_{13}(t) & \rho^*_{23}(t) & \rho_{33}(t) & \rho_{34}(t) \\
\rho^*_{14}(t) & \rho^*_{24}(t) & \rho^*_{34}(t) & \rho_{44}(t) \\
\end{pmatrix}
\ee

Defining the effective rate parameters $\Gamma_R =\frac{1}{\hbar}\, \left(   |\alpha_R^{(1)}|^2 + |\alpha_R^{(2)}|^2 \right )  \,\rho_R$,  $\Gamma_L =\frac{1}{\hbar}\, \left(   |\alpha_L^{(1)}|^2 + |\alpha_L^{(2)}|^2 \right )  \,\rho_R$, the time evolution of the density matrix is given by:
\ba  \label{Masters1}
\dot{\rho}_{11}(t)&=&-\frac{\rho_{11}(t)\alpha_R^2}{1+e^{-2 \beta
  t_{34}}}+\frac{\Gamma_R
   \rho_{33}(t)}{1+e^{2 \beta t_{34}}}-\frac{\Gamma_L \rho_{11}(t)}{1+e^{-2 \beta t_{12} (t)}}+\frac{\Gamma_L \rho_{44}(t)}{1+e^{2
   \beta t_{12}(t)}}  +2 t_{23} \text{Im}(\rho_{12}(t))\n
\dot{\rho}_{22}(t)&=&\frac{\rho_{44}(t)
  \Gamma_R}{1+e^{-2 \beta t_{34}}}-\frac{\Gamma_R \rho_{22}(t)}{1+e^{2 \beta
  t_{34}}}+\frac{\Gamma_L \rho_{33}(t)}{1+e^{-2 \beta t_{12} (t)}}-\frac{\Gamma_L \rho_{22}(t)}{1+e^{2 \beta t_{12} (t)}}-2 t_{23}
   \text{Im}(\rho_{12}(t))\n
\dot{\rho}_{33}(t)&=&\frac{\rho_{11}(t)\Gamma_R}{1+e^{-2
   \beta t_{34}}}-\frac{\Gamma_R \rho_{33}(t)}{1+e^{2 \beta
  t_{34}}}-\frac{\Gamma_L \rho_{33}(t)}{1+e^{-2 \beta t_{12} (t)}}+\frac{\Gamma_L \rho_{22}(t)}{1+e^{2 \beta t_{12}(t)}}+2 t_{23} \text{Im}(\rho_{34}(t))\n
\dot{\rho}_{44}(t)&=&-\frac{\rho_{44}(t)
  \Gamma_R}{1+e^{-2 \beta t_{34}}}+\frac{\Gamma_R \rho_{22}(t)}{1+e^{2 \beta
  t_{34}}}+\frac{\Gamma_L \rho_{11}(t)}{1+e^{-2
   \beta t_{12}(t)}}-\frac{\Gamma_L
   \rho_{44}(t)}{1+e^{2 \beta t_{12}(t)}}-2  t_{23} \text{Im} (\rho_{34}(t)) \n
\dot{\rho}_{12}(t)&=&\frac{1}{2} \left(\rho_{12}(t)
   \left(-\frac{\Gamma_R}{1+e^{-2 \beta t_{34}}}-\frac{\Gamma_R}{1+e^{2
   \beta t_{34}}}-\frac{\Gamma_L}{1+e^{-2 \beta t_{12} (t)}}-\frac{\Gamma_L}{1+e^{2 \beta t_{12}(t)}}+4 it_{34}+4 i
  t_{12}(t)\right)-2 it_{23}
   (\rho_{11}(t)-\rho_{22}(t))\right)\n
\dot{\rho}_{21}(t)&=&\frac{1}{2} \left(\rho_{21}(t) \left(-\frac{\Gamma_R}{1+e^{-2 \beta
  t_{34}}}-\frac{\Gamma_R}{1+e^{2 \beta
  t_{34}}}-\frac{\Gamma_L}{1+e^{-2 \beta t_{12}(t)}}-\frac{\Gamma_L}{1+e^{2 \beta t_{12}(t)}}-4 it_{34}-4 i
  t_{12}(t)\right)+2 it_{23}
   (\rho_{11}(t)-\rho_{22}(t))\right)\n
\dot{\rho}_{34}(t)&=&\frac{1}{2}
   \left(\rho_{34}(t) \left(-\frac{\Gamma_R}{1+e^{-2 \beta
  t_{34}}}-\frac{\Gamma_R}{1+e^{2 \beta
  t_{34}}}-\frac{\Gamma_L}{1+e^{-2 \beta t_{12} (t)}}-\frac{\Gamma_L}{1+e^{2 \beta t_{12}(t)}}-4 it_{34}+4 i
  t_{12}(t)\right)-2 it_{23}
   (\rho_{33}(t)-\rho_{44}(t))\right)\n
\dot{\rho}_{43}(t)&=&\frac{1}{2}
   \left( \rho_{43}(t) \left(-\frac{\Gamma_R}{1+e^{-2 \beta
  t_{34}}}-\frac{\Gamma_R}{1+e^{2 \beta
  t_{34}}}-\frac{\Gamma_L}{1+e^{-2 \beta t_{12}(t)}}-\frac{\Gamma_L}{1+e^{2 \beta t_{12}(t)}}+4 it_{34}-4 i
  t_{12}(t)\right)+2 it_{23}
   (\rho_{33}(t)-\rho_{44}(t))\right)\n
   \ea

   \ba \label{Masters2}
\dot{\rho}_{13}(t)&=&\frac{\rho_{42}(t)
  \Gamma_L}{1+e^{2 \beta t_{12}(t)}}+\left(-\frac{\Gamma_R}{2
   \left(1+e^{-2 \beta t_{34}}\right)}-\frac{\Gamma_R}{2 \left(1+e^{2 \beta
  t_{34}}\right)}-\frac{\Gamma_L}{1+e^{-2 \beta t_{12}(t)}}+2 i
  t_{34}\right) \rho_{13}(t)-it_{23}
   (\rho_{14}(t)-\rho_{23}(t))\n
\dot{\rho}_{31}(t)&=&\frac{\rho_{24}(t)
  \Gamma_L}{1+e^{2 \beta t_{12}(t)}}+\left(-\frac{\Gamma_R}{2
   \left(1+e^{-2 \beta t_{34}}\right)}-\frac{\Gamma_R}{2 \left(1+e^{2 \beta
  t_{34}}\right)}-\frac{\Gamma_L}{1+e^{-2 \beta t_{12}(t)}}-2 i
  t_{34}\right) \rho_{31}(t)-it_{23}
   (\rho_{32}(t)-\rho_{41}(t))\n
\dot{\rho}_{24}(t)&=&\frac{\rho_{31}(t)\Gamma_L}{1+e^{-2 \beta
  t_{12}(t)}}+\left(-\frac{\Gamma_R}{2 \left(1+e^{-2 \beta
  t_{34}}\right)}-\frac{\Gamma_R}{2 \left(1+e^{2 \beta
  t_{34}}\right)}-\frac{\Gamma_L}{1+e^{2 \beta t_{12}(t)}}-2 i
  t_{34}\right) \rho_{24}(t) +it_{23}
   (\rho_{14}(t)-\rho_{23}(t)) \n
\dot{\rho}_{42}(t)&=&\frac{\rho_{13}(t)\Gamma_L}{1+e^{-2
   \beta t_{12}(t)}}+\left(-\frac{\Gamma_R}{2 \left(1+e^{-2 \beta
  t_{34}}\right)}-\frac{\Gamma_R}{2 \left(1+e^{2 \beta
  t_{34}}\right)}-\frac{\Gamma_L}{1+e^{2 \beta t_{12}(t)}}+2 i
  t_{34}\right) \rho_{42}(t) +it_{23}
   (\rho_{32}(t)-\rho_{41}(t)) \n
\dot{\rho}_{14}(t)&=&-\frac{\rho_{32}(t)\Gamma_R}{1+e^{2
   \beta t_{34}}}+
   \left(-\frac{\Gamma_R}{1+e^{-2 \beta t_{34}}}-\frac{\Gamma_L}{2
   \left(1+e^{-2 \beta t_{12}(t)}\right)}-\frac{\Gamma_L}{2
   \left(1+e^{2 \beta t_{12}(t)}\right)}+2 it_{12} (t)\right)\rho_{14}(t)-it_{23} (\rho_{13}(t)-\rho_{24}(t))\n
\dot{\rho}_{41}(t)&=&-\frac{\rho_{23}(t)
  \Gamma_R}{1+e^{2 \beta t_{34}}}+
   \left(-\frac{\Gamma_R}{1+e^{-2 \beta t_{34}}}-\frac{\Gamma_L}{2
   \left(1+e^{-2 \beta t_{12}(t)}\right)}-\frac{\Gamma_L}{2
   \left(1+e^{2 \beta t_{12}(t)}\right)}-2 it_{12} (t)\right)\rho_{41}(t)+it_{23}
   (\rho_{31}(t)-\rho_{42}(t))\n
\dot{\rho}_{23}(t)&=&-\frac{\rho_{41}(t)\Gamma_R}{1+e^{-2 \beta t_{34}}}+
   \left(-\frac{\Gamma_R}{1+e^{2 \beta
  t_{34}}}-\frac{\Gamma_L}{2 \left(1+e^{-2 \beta t_{12} (t)}\right)}-\frac{\Gamma_L}{2 \left(1+e^{2 \beta t_{12} (t)}\right)}-2 it_{12}(t)\right)\rho_{23}(t)+i
  t_{23} (\rho_{13}(t)-\rho_{24}(t))\n
\dot{\rho}_{32}(t)&=&-\frac{\rho_{14}(t)\Gamma_R}{1+e^{-2
   \beta t_{34}}}+\left(-\frac{\Gamma_R}{1+e^{2 \beta
  t_{34}}}-\frac{\Gamma_L}{2 \left(1+e^{-2 \beta t_{12} (t)}\right)}-\frac{\Gamma_L}{2 \left(1+e^{2 \beta t_{12} (t)}\right)}+2 it_{12}(t)\right)\rho_{32}(t) -it_{23}
   (\rho_{31}(t)-\rho_{42}(t))\n
   \ea

\end{widetext}
where $\rho_{ji } = \rho^*_{ij}$.   We see that the 16 components of the density matrix can be divided into two blocks: the elements in the $2 \times 2$ block diagonals (whose time evolution is given in Eq. (\ref{Masters1}), and the $ 2 \times 2 $ block off-diagonals (Eq. \ref{Masters2}).  If the system begins either in thermal equilibrium (where the density matrix is diagonal) or in an eigenstate of the Majorana hopping Hamiltonian, all elements in the block off-diagonals are initially 0, and hence these elements remain 0 throughout the time evolution.  Hence we can focus our attention on the 8 block diagonal elements, whose time evolution is given in Eq. (\ref{Masters1}).

First, let us determine the equilibrium solutions.  If $t_{12}$ is time-independent, these are given by $\rho_{ij} =0, i\neq j$, and
\ba
\rho_{11}& =& n_F(2 t_{12} )   n_F(2 t_{34} )\n
\rho_{22} & =& n_F(2 t_{12} )   n_F(2 t_{34} )e^{ 2 \beta (t_{12}+t_{34}) } \n
\rho_{33} & =&n_F(2 t_{12} )   n_F(2 t_{34} )e^{ 2 \beta t_{34} }\n
\rho_{44} & =&n_F(2 t_{12} )   n_F(2 t_{34} )e^{ 2 \beta t_{12} }
\ea
At high temperatures, this simply indicates that the probability of finding the system in any of the four states is $1/4$; at low temperatures the relative probability of being found in each state is weighted by its energy, as expected.    If $t_{12}$ is changing in time, at low temperatures there is no equilibrium solution.
There is however a quasi-stationary solution which oscillates with the driving frequency and is phase locked with the driving.

To study the time-dependent equations analytically, we take the system to be at infinite temperature. Defining
\ba
 \Gamma  =  \frac{\Gamma_R+\Gamma_L}{2}  \ \ \ \ \
  \Gamma_A  =  \frac{\Gamma_R-\Gamma_L}{2}  \n
  \rho_{e,s} =\rho_{11} + \rho_{22} \ \ \ \ \    \rho_{e,a} =\rho_{11} - \rho_{22} \n
   \rho_{o,s} =\rho_{33} + \rho_{44} \ \ \ \ \    \rho_{o,a} =\rho_{33} - \rho_{44} \n
 x_{12} = \text{Re} \left( \rho_{12} \right )  \ \ \ \ \ y_{12} = \text{Im} \left( \rho_{12}\right ) \n
  x_{34} = \text{Re}\left(  \rho_{34}\right ) \ \ \ \ \ y_{34} = \text{Im}\left(  \rho_{34} \right )
\ea
we find that the equations of motion for the symmetric components $  \rho_{e,s},   \rho_{o,s}$ of the diagonal elements of the density matrix are:
   \ba  \label{Mds}
\dot{\rho}_{e,s}(t)&=& - \Gamma \left( \rho_{e,s}(t) -  \rho_{o,s}(t) \right ) \n
\dot{\rho}_{o,s }(t)&=&  - \Gamma \left( \rho_{o,s}(t) -  \rho_{e,s}(t) \right )
 \ea
 This indicates that $ d/dt \text{Tr}( \rho) =0$, and that the difference in the total probabilities of finding the system in the sector of even versus odd fermion parity decays exponentially to $1/2$.
 The equations of motion for the remaining six components of $\rho$ are:
 \ba \label{DiffEqs}
\dot{\rho}_{e,a}(t)&=&   - \Gamma \rho_{e,a}(t) + \Gamma_A \rho_{o,a}  - 4 t_{23} y_{12}(t) \n
\dot{\rho}_{o,a}(t)&=& - \Gamma \rho_{o,a}(t) + \Gamma_A \rho_{e,a}  - 4  t_{23} y_{34}(t) \n
\dot{x}_{12}(t)&=& - \Gamma x_{12}(t)  + 2  \left( t_{34}+ t_{12}(t)  \right)y_{12}(t) \n
 \dot{y}_{12}(t)&=&-  \Gamma y_{12}(t)  - 2  \left( t_{34}+ t_{12}(t)  \right)x_{12}(t) + t_{23} \rho_{e,a} \n
 \dot{ x}_{34}(t)&=& - \Gamma x_{34}(t)  - 2  \left( t_{34}- t_{12}(t)  \right)y_{34}(t) \n
 \dot{y}_{34}(t)&=&-  \Gamma y_{34}(t)  + 2  \left( t_{34}- t_{12}(t)  \right)x_{34}(t) + t_{23} \rho_{o,a} \n
    \ea
If the effective rate constants for the right and left wire segments are equal, then $\Gamma_A =0$ and this reduces to two sets of three coupled first order differential equations.
For $t_{12}$ independent of time, these can be solved analytically to give:
\begin{widetext}
\ba \label{Tsols1}
\rho_{e,a} (t)\ &=&  \frac{ e^{-\Gamma t} }{\epsilon_e^2}\left(\rho_{e,a}(0)\left(
   t_{23}^2 \cos (2 \epsilon_e t)+ t_{12,34}^2\right)+2 x_{12}(0) t_{23}  t_{12,34}  \left( 1 -\cos (2 \epsilon_e t) \right)
   -2 y_{12}(0) \epsilon_e t_{23} \sin (2 \epsilon_e t) \right )
  \n
 x_{12}(t)&=& \frac{e^{-\Gamma t} }{2  \epsilon_e^2}
   \left( 2 x_{12}(0) \left(  t_{23}^2 - t_{12,34} ^2\cos (2 \epsilon_e t) \right ) +\rho_{e,a}(0)
   t_{23} t_{12,34} \left(  1 - \cos (2 \epsilon_e t) \right)+2 y_{12}(0) \epsilon_e t_{12,34} \sin (2 \epsilon_e
   t)  \right)\n
y_{12}(t)&=& \frac{e^{-\Gamma t} }{2 \epsilon_e}\left ( 2 y_{12}(0) \epsilon_e \cos (2
   \epsilon_e t) - 2 x_{12}(0)
   t_{12,34} \sin (2 \epsilon_e t) +\rho_{e,a}(0) t_{23} \sin (2 \epsilon_e t)+\right)
\ea
\end{widetext}
where $t_{12,34} = t_{12} + t_{34}$, and $\epsilon_e = \sqrt{ (t_{12} + t_{34})^2 + t_{23}^2 }$ is the modulus of the band energy in the even-parity sector.   (The form of the solution is similar in the odd sector).
We are principally interested in the behavior of the diagonal component $\rho_{e,a}(t)$, which has oscillations only for $t_{23} \neq 0$.  These occur because the states $|0,0 \rangle$ and $|1,1 \rangle$ are not exact eigenstates.  All oscillations here are at the scale of the band gap in the even-parity sector.

Next, we consider solutions where $t_{12}$ is of the form
\be
t_{12} = t_{12}^{(0)} \cos \omega_0 t
\ee
When $t_{12}$ is oscillating, analytic solutions are more complicated but the equations are easily solved numerically.  However, if $t_{12}$ varies slowly compared to $\epsilon_e$, the dynamics described by Eq. (\ref{Masters1}) is approximately given by Eq. (\ref{Tsols1}).   Thus we can obtain a good approximate description of the true dynamics by assuming that Eq. (\ref{Tsols1}) describes the evolution everywhere except in the vicinity of the avoided crossing, where Eq. (\ref{Masters1}) also includes the probability of Landau-Zener tunneling.
The probability of tunneling out of the ground state at the crossing is approximately
\be \label{LZTunnels}
P_T = e^{ -   \tau} \ ,\  \ \ \ \
\tau = \frac{\pi t_{23}^2}{  \hbar \omega_0 t_{12}^{(0)} \cos^{-1} (-t_{34}/ t_{12}^{(0)}  ) }
\ee
for every sweep past the avoided crossing.  (Here $\omega_0 t_{12}^{(0)}\cos^{-1} (-t_{34}/ t_{12}^{(0)}  )$ is the rate at which the energy splitting between the two bands is shrinking in the vicinity of the avoided crossing ).
If $P_T$ is small (a.k.a. if $t_{23}^2$ is large relative to the rate of change of the relative energies) then the system remains in its ground state with high probability -- meaning that at the avoided crossing, it changes between an eigenstate that is predominantly $|1,1 \rangle$ to one that is predominantly $|0,0 \rangle$.  If $P_T$ is large, the system is likely to end up in an excited state, in which case it will be predominantly $|1,1 \rangle$ (or $|0,0\rangle$) on both sides of the avoided crossing.

Landau-Zener tunneling leaves $\rho_{e, s}(t)$ (the total probability to be in the even sector) invariant.  Thus we can approximate its effect by applying the following transformation to $\rho_{e,a} $ each time the system passes the avoided crossing:
%\be
%\rho_{11}(t) \rightarrow P_T \rho_{11}(t) + (1- P_T) \rho_{22}(t)
\be
\rho_{e,a} (t) \stackrel{t=t_{\text{cross}} }{\rightarrow} - ( 1 - 2 P_T) \rho_{e,a}(t)
\ee
Hence for $P_T < 1/2$, $\rho_{e,a}$ changes sign twice per cycle.   At each sign change there is also a decrease in the magnitude of $\rho_{e,a}$, which depends on the size of $P_T$.  If $P_T > 1/2$,  $\rho_{e,a}$ is always of the same sign (the system remains dominated by $|1,1\rangle$ or $|0,0\rangle$ components), but decreases in magnitude at each cycle by an amount that depends on $(1- P_T)$.

This approximate description of the dynamics (Eq. \ref{Masters1}) indicates that there are two time-scales for oscillations in the problem: there are fast oscillations at frequency $2 \epsilon_e$, of amplitude $t_{23}^2 \rho_{e,a}$, %($2 \epsilon_o$) in the even (odd) sector,
and slow oscillations at a frequency set by the driving frequency $\omega_0/2$, of amplitude $(1- 2 P_T) \rho_{e,a}$.   (For the approximate description in the preceding paragraphs to be valid, the
``slow" oscillations must indeed be slow relative to $2 \epsilon_e$; otherwise there is a significant probability of Landau-Zener tunneling away from the avoided crossing, leading to a different dynamical regime for Eq. (\ref{DiffEqs}).)  Importantly, if $t_{23}$ is small relative to the other hoppings, the amplitude of the fast oscillations is small (for appropriate initial conditions) relative to the constant term in Eq. (\ref{Tsols1}), and these are easily distinguished from the oscillations that arise from Landau-Zener tunneling for modest values of $P_T$.  This can be clearly seen in the numerical results of Fig. \ref{RhoPlots}.

%(Because the avoided crossing occurs at $t_{12} = - t_{34}$ ($t_{34}$) in the even (odd) sector, these oscillations occur twice in each cycle of $t_{12}$, at times
%\be
%t= \frac{1}{\omega_0} \cos^{-1} (t_{34}/ t_{12}^{(0)}  ), t= \frac{1}{\omega_0} \left( 2 \pi - \cos^{-1} (-t_{34}/ t_{12}^{(0)}  ) \right )
%\ee
%).

%
%\section{Measuring correlations from the density matrix time evolution}
%\label{rhoMeasApp}
%
%{\bf SASHA: Fiona, I removed this chapter. The Latex code is at the end of the file.}

\bibliography{ParityBib,library}

\end{document}